# Connection between forest fire emission and COVID-19 incidents in West Coast regions of the United States

*Srikanta Sannigrahi[a*], Arabinda Maiti[b], Francesco Pilla[a], Qi Zhang[c], Somnath Bar[d], Saskia Keesstra[e, f], Artemi Cerda[g]*

[a] School of Architecture, Planning and Environmental Policy, University College Dublin Richview, Clonskeagh, Dublin, D14 E099, Ireland.

[b] Department of Geography, Vidyasagar University, Midnapore, West Bengal, India

[c] Department of Earth & Environment and Frederick S. Pardee Center for the Study of the Longer-Range Future, Boston University, Boston, MA 02215, USA

[d] Department of Geoinformatics, Central University of Jharkhand, Ranchi, India

[e] Team Soil, Water and Land Use, Wageningen Environmental Research, Wageningen University & Research, Wageningen, Netherlands

[f] Civil, Surveying and Environmental Engineering and Centre for Water Security and Environmental Sustainability, The University of Newcastle, Callaghan 2308, Australia

[g] Soil Erosion and Degradation Research Group, Department of Geography, Valencia University, Blasco Ibàñez, 28, 46010 Valencia, Spain

*Corresponding author: **Srikanta Sannigrahi**

E-mail: (Srikanta Sannigrahi*) : srikanta.sannigrahi@ucd.ie

## Abstract

Forest fires impact on soil, water and biota resources has been widely researched. Although forest fires profoundly impact the atmosphere and air quality across the ecosystems, much less research has been developed to examine its impact on the current pandemic. The recent West Coast forest fire in the United States (US) caused severe environmental and public health burdens. As of October 21, nearly 8.2 million acres of forest area were burned, and more than 25 casualties were reported so far. In-situ air pollution data were utilized to examine the effects of 2020 forest fire on atmosphere and coronavirus (COVID-19) casualties. The spatial-temporal concentrations of particulate matter ($PM_{2.5}$ and $PM_{10}$) and Nitrogen Dioxide ($NO_2$) were collected from August 1 to October 30 for 2020 (fire year) and 2019 (reference year). Both spatial (Multiscale Geographically Weighted Regression) and non-spatial (negative binomial regression) regression analysis was performed to assess the adverse effects of fire emission on human health. The in-situ data-led measurements showed that the maximum increases in $PM_{2.5}$, $PM_{10}$, and $NO_2$ concentrations ($\mu g/m^3$) were clustered in the West Coastal fire-prone states during the August 1 – October 30 period. The average concentration ($\mu g/m^3$) of particulate matter ($PM_{2.5}$ and $PM_{10}$) and $NO_2$ was increased in all the fire states affected badly by forest fires. The average $PM_{2.5}$ concentration ($\mu g/m^3$) over the period was recorded as 7.9, 6.3, 5.5, and 5.2 for California, Colorado, Oregon, and Washington in 2019, which was increased up to 24.9, 13.4, 25, and 17 in 2020. Both spatial and non-spatial regression models exhibited a statistically significant association between fire emission and COVID-19 incidents. A total of 30 models were developed for analyzing the spatial non-stationary and local association between the predictor and response factors. All these spatial models have demonstrated a statistical significant association between fire emissions and COVID counts. More thorough research is needed to better understand the complex association between forest fire and human health.

## 1. Introduction

Forest fire is now becoming an increasing global environmental threat across the ecosystem and caused severe public health burdens due to the upsurges of smokes and particulate matter concentration into the lower atmosphere (B. et al., 2011; Bowman and Johnston, 2005; Fowler, 2003; Goldammer et al., 2008). Amongst the causal factors, the climate change and associated factors (rising temperature, long dry spell, lack of soil moisture, the abundance of flammable materials, etc.) have augmented the severity, intensity, and length of forest fire season and eventually increases the exposure to hazardous air pollutants in the area under forest fire threats (Aponte et al., 2016; Flannigan et al., 2000; Mateus and Fernandes, 2014; Meira Castro et al., 2020; Michel Arbez et al., 2001). Forest fire has been significantly associated with the increases in gaseous, i.e. Carbon (Aragão et al., 2018; Lazaridis et al., 2008; Lü et al., 2006), Smoke (Fromm and Servranckx, 2003; Johnston et al., 2014; Mott et al., 2002), black Carbon (Badarinath et al., 2007; Jeong et al., 2004), aerosol (Pio et al., 2008; Randerson et al., 2006), fine (Ikemori et al., 2015; McLean et al., 2015; Sapkota et al., 2005), coarse particulate matter (Henderson et al., 2008; Juneng et al., 2009), Nitrogen oxides (McEachern et al., 2000; Spichtinger et al., 2001), and other pollutants (NO, $O_3$, VOC) (Cheng et al., 1998). In Greece, forest fire emission was found to be the most significant contributor to the air pollution problem during the fire occurrence period (Cheng et al., 1998). Through long and short-range atmospheric transport, forest fire emissions impacted a large region from its source (Lazaridis et al., 2008). The effects of forest fire emission on air pollution could be easily detected in rural areas where anthropogenic emission is limited (Cheng et al., 1998). In the rural region of Edmonton (Canada), the hourly $NO_x$, $O_3$ concentration was recorded 50–150% higher than the seasonal median values, which can be attributed to the forest fire and resulted in emission (Cheng et al., 1998). However, the forest fire emitted pollutants can travel thousands of kilometres with the help of upper atmospheric circulation and exacerbate the problem of local air pollution in the heavily polluted regions (Sapkota et al., 2005). The deterioration of air quality in Baltimore city (located nearly 1100 km away from the fire source region) due to 2002 Canadian forest fires has once again proven the fact that forest fire emission is not only posing threats to the nearby communities, but the same can have substantial public health impact to the regions located far away from the fire-affected areas (Sapkota et al., 2005).

Several earlier research has utilized many available resources, i.e. satellite estimates (Konovalov et al., 2011; Wu et al., 2006), in-situ measurement (Konovalov et al., 2011), low-cost sensor measurements (Delp and Singer, 2020; Sayahi et al., 2019), air pollution models (Watson et al., 2019) to examine the detrimental impact of forest fires on air quality across the ecosystems. Wu et al. (2006) noted that the concentration of $PM_{10}$ was increased up to 160 $\mu g/m^3$ due to the 2003 southern California forest fires and resulted in the emission of particulate matter. (Hodzic et al., 2007) study on 2003 European forest fires documented a drastic increase (20 to 200%) of air pollutants, especially $PM_{10}$, due to the emission of gaseous compounds during the fire period. Forest fires were also found to be highly associated with the increases of fine particulate matter ($< _{2.5}\mu m$) (Jaffe et al., 2008; Matz et al., 2020; Sullivan et al., 2008). Though there has been strong and clear evidence that forest fires have a strong negative impact on air quality, still, there are several other confounding factors, such as the description of fire emissions, atmospheric dispersion of smoke, and the chemical transformations of smoke, etc., needs to be evaluated comprehensively in order to understand the association between forest fire and air quality in a better way (Martins et al., 2012).

Several studies have reported the association between short/long term exposure to air pollution and incidents of severe acute respiratory syndrome coronavirus 2 - 2019 (SARS-CoV-2 – COVID-19) in many regions across the world (Ogen, 2020; Sciomer et al., 2020; Shen et al., 2020; Wang et al., 2020; Zhu et al., 2020). Amongst the key air pollutants, the concentration of particulate matter ($PM_{2.5}$ and $PM_{10}$) and its association with COVID casualties has been the central focus in these studies. Zhu et al. (2020) study has performed Generalized Additive Model (GAM) after considering COVID-19 incidences of 120 cities in China and found that a10 µg/m$^3$ increases $PM_{2.5}$ and $PM_{10}$ was associated with a 2.24% and 1.76% increase in daily COVID-19 confirmed cases. Zhu et al. found that per 10 µg/m$^3$ increase of $NO_2$ was associated with a 6.94% increase in daily COVID-19 confirmed cases. (Yao et al., 2020) study analyzed the linkages between air pollution and COVID incidences in 49 cities in China using a multiple linear regression model and found that per 10 µg/m$^3$ increase in $PM_{2.5}$ and $PM_{10}$ was associated with a 0.24% (0.01% - 0.48%) and 0.26% (0.00% - 0.51%) increase in the daily COVID-19 fatality rate. (Wu et al., 2020) observation considered 3000 counties of the USA and performed a zero-inflated negative binomial model to examine the linkages between the concentration of $PM_{2.5}$ and COVID death rate during the January to April 4, 2020 period. They have reported that a 1 µg/m$^3$ long-term exposure increase in $PM_{2.5}$ was associated with a 15% increase in COVID-19 death rate. In England, (Travaglio et al., 2021) study had found a strong association between $PM_{2.5}$ concentration and COVID incidents (an increase of 1 µg/m$^3$ in the long-term average of $PM_{2.5}$ was associated with a 12% increase in COVID-19 cases).

The unprecedented and record-breaking forest fire events in 2020 in the West Coast states (California, Oregon, Washington, Colorado) in the USA can cause severe health burdens, especially at the time of the COVID pandemic. As of October 21, 2020, nearly 8.2 million acres (33,000 km$^2$) of forest area were burnt, and 46 casualties have been reported so far (https://www.fire.ca.gov/). The economic cost attributed to these events can be as much as $2.707 Billion in 2020 unit price. The air quality during the fire periods has become extremely poor and reached very unhealthy to hazardous level in many fire-affected regions. Though the air quality has improved in some parts of the area due to lockdown measures, it still remains hazardous in the areas poorly affected by the forest fire. Since it has been proven that forest fire contributes substantially in adding gaseous and particulate matter concentration into the lower and upper atmosphere, the coexistence of two extreme events, i.e. the 2020 forest fire, which has declared as the most intense forest fire in the USA since 2003, and COVID-19 pandemic, which is also announced as one of the worst pandemics in the history of human civilization, has not been discussed thoroughly. Therefore, the synergistic association between these two rare events (West Coast forest fire and COVID casualties) and their combined effects on human health should be examined so that the same would allow us to understand how climate change led extremities can exacerbate the crisis of public health. The objectives of this are: (1) examine the air quality levels during the fire period (August 01 to October 30) in fire year (2020) and reference year (2019), (2) measures the changes in air quality due to forest fire; (3) analzying the association between fire emission and COVID incidences using spatial regression models.

## 2. Materials and methods:

## 2.1 Data source and processing

The Western states of the USA, mainly California, Oregon, Washington, Idaho, Colorado, have witnessed the record-breaking (surpassed the last 18 years fire severity records in terms of forest area burnt and damages of property and structures) forest fires that started early in August 2020. As of Mid October 2020, 8.2 million acres (33,000 square kilometres) were burned, and at least 46 casualties have been reported so far (National Interagency Fire Center). Many factors, including climate change, led to an extremely higher temperature, lack of surface moisture due to below-normal precipitation in the preceding seasons, extended dry spell and associated heat waves, higher wind speed, abundant fuel load, etc. have exacerbated the severity and length of the forest fires in these regions (Balch et al., 2017; Collins et al., 2019; Kane et al., 2017).

In-situ air pollution measurements were retrieved from OpenAQ[1]. OpenAQ is a non-profit organization aiming to retrieve, harmonize, share open-air quality information to citizens and organizations, and provide an up-to-date status of clean air, which would eventually help prevent air pollution led health burdens across the world. The OpenAQ platform retrieved the latest and up-to-date air quality data from multiple sources such as government/institution air quality monitoring stations and low-cost open-air quality sensors. All the data were collected from August 1 to October 30 period for comparative assessment and subsequent interpretation.

Daily COVID-19 cases and death data were collected from USAFacts[2]. The USAFacts collects COVID counts data from the Center for Disease Control and Prevention (CDC), State and local-level public health agencies. The county-level data for each States collects and verified with the local and State agencies. The daily cumulative sum of deaths and cases reports for each administrative units were recorded through manual entry or web scraping (USAFacts, 2020). For California and Texas, USAFacts gather COVID data from each county's public health website. Currently, the presumptive positive cases considered as confirmed cases, which is in line with CDC's COVID reporting. Daily and cumulative sum counts of cases and deaths were analyzed to find its association with forest fire emissions. More details about the COVID data collection process, quality assurance, data collection assumptions, flag detection and reporting, etc., can be found on the USAFacts website.

The fire emission was measured for the study region using Global Fire Emission Database Explorer v4.1 (GFEDv4s[3]) (Akagi et al., 2011; Andreae and Merlet, 2001; Giglio et al., 2013; Randerson et al., 2012; van der Werf et al., 2017, 2017). Time series GFED data was incorporated into the assessment to estimate particulate matter and greenhouse gas emissions from forest fires. GFED data has contained 1440 columns and 720 rows with 0.25° spatial resolution and available from 1997 to the current date. In this study, GFED estimates were utilized for the 1997 – 2020 period. GFED emission data comes with a three-time scale, i.e. annual emissions, monthly emission, and daily emission of gaseous and particulate matter components. Each raster layer consists of three main datasets: the spatial extent of the burned area, monthly emissions and fractional contributions of different fire types, and daily / 3-hourly emission records at a specified spatial scale. Since the earlier version of the GFED has not considered the small fire details, the present study utilized the most updated GFED4.1s (with small fire) statistics for analysis and subsequent interpretation. Both emissions and burned area information were used to calculate fire-led emission for the 1997 to 2020 period. Fires that have been recorded over varied landscapes also considered for the analysis. GFED4.1s offers detailed statistics of LULC specific small fire information. Among the regions, fire emissions data for the Temperate North America (TENA) regions were considered for this analysis.

---

[3] http://www.globalfiredata.org/index.html

## 2.2 Regression analysis

### 2.2.1 Non-spatial regression model

Two regression models, i.e. Ordinary Least Square (OLS) and Negative Binomial Regression was performed to examine the association between forest fire emission and COVID incidences at the county scale. For regression analysis, only fire States were considered. OLS measures the interaction and association between the sets of dependent and independent factors (Maiti et al., 2021; Mollalo et al., 2020; Oshan et al., 2019; Sannigrahi et al., 2020b, 2020a). Additionally, OLS fits a line based on the characteristics of the dependent and independent observations in the bivariate data framework to minimize the squared distance of each data points from the fitted line (Kilmer and Rodríguez, 2017; Sokal et al., 1995). The OLS can be formed as follows

$$y_i = a + \beta x_i + \varepsilon_i \qquad (1)$$

Where $a$ is the intercept, $\beta$ vector of regression coefficients, $xi$ is the vector of selected air pollutants at county $i$, $\varepsilon_i$ is the error term.

In addition to the standard OLS estimates, the negative binomial (NB) regression method was also applied for analyzing the association between forest fire emission and associated COVID case/death counts during the fire period (Copat et al., 2020; Wu et al., 2020). The NB regression model was used in the present research to minimize overdispersion in statistical estimates. Among the several link function, the log has been used as a link function to carry out the NB regression analysis. The NB regression with Poisson parameter can be expressed as follows:

$$\iota_n(\lambda_i) = \beta_0 + \sum_{k=1}^{p} \beta_k X_{ik} + \theta_i \qquad (2)$$

Where the Poisson parameter ($\lambda_i$) represents the expected estimates of observation i, $\beta_0$, $\beta_k$ are the model input parameters, $X_{ik}$ is the kth number of the independent variable for observation i, $\theta_i$ is the gamma-distributed error (*e*) term with mean and variance considered as 1 and α (Chang, 2005; Wang et al., 2021).

### *2.2.2 Spatial regression model*

In the present study, a modified version of Geographically weighted regression (GWR) – MGWR, developed by Oshan et al. (2019), has been used for exploring the spatially varying association between the fire emission and COVID counts at county scale in the contiguous US. The GWR is a local spatial regression approach conceptualized upon the assumption of spatial heterogeneity and spatial non-stationarity among the parameters in a feature space. Unlike global regression, which assumes a spatial homogeneity and constant relationship among the features, the GWR often increases the model fit by reducing residuals of spatial autocorrelation in parameter estimates. GWR is also sensitive to bandwidth and kernel selection and parametrization, which seeks special attention while designing the spatial models to explain any spatial varying relationship between parameters.

Recently, an extension to the existing GWR frameworks known as multiscale GWR (MGWR) allows exploring the locally varying association between the parameters at a unique spatial scale which eventually helps to understand the multiscale analysis of spatial heterogeneity and spatial non-stationarity. MGWR that eliminates the assumption and limitation of the existing GWR that all spatial association vary uniformly at all spatial scale could reduce the model overfitting, spatial auto-correlation and uncertainty that mainly originates from the scale-dependent approximation of parameters. In the present study, the MGWR model incorporated into the spatial analysis framework to explore the spatial association between forest fire emission and COVID counts at the lowest spatial in the contiguous USA. The mgwr python package, which has many dependencies, such as NumPy, SciPy, pandas, matplotlib, libpysal, and spglm, was used to run the MGWR model. A standard GWR model can be expressed as:

$$y_i = \beta_{i0} + \sum_{k=1}^{\rho} \beta_{ik} x_{ik} + \varepsilon_i$$

(3)

Where $y_i$ is the dependent variable (COVID case/death in this study) for location $i$, $\beta_{i0}$ is the intercept coefficient of location i, $\beta_{ik}$ is the local regression coefficient of $k$ th explanatory variable at location i, $x_{ik}$ is the kth independent variable at location i, $\varepsilon_i$ is the random error term at location i.

In the MGWR model, the scale or bandwidth dependent spatial assumption in the GWR model has been respecified by incorporating scale variant association between the parameters in feature space as follows:

$$y = \sum_{j=1}^{k} f_j + \varepsilon_i$$

(4)

Where $f_j$ is the smoothing function introduced in MGWR model, which applied to the $j$th independent variable and allowed to explore the association at distinct spatial scale/bandwidth.

**2.3 Experimental Design**

Different approaches have been adopted in each segment of the analysis, which collectively demonstrates how forest fire emission can cause severe real-time and lagged impact on the atmosphere and public health. The entire analysis was performed through many successive steps: *first*, the fire affected States, i.e. California, Oregon, Washington, and Colorado, were demarcated based on the number of active fire events statistics and National Fire Emergency Report. These States have been grouped together and named as fire States (short name used as FireStates). The remaining States assumed to have fewer fire effects have been marked as ExFireStates (Excluding fire States). Accordingly, the air pollution assessment was performed for the (a) fire States (FireSates), (b) ExFireStates, and (c) all States (considered both FireStates and ExFireStates), from August 1 to October 30 in both reference year (2019) and fire year (2020). 2019 was chosen as a reference year as this year has comparably lower active fire events than the preceding years. Moreover, OpenAQ air pollution data is available from 2018; this could be another reason for choosing 2019 as a reference year and 2020 as a fire year. 2018 has not been considered in the analysis as there had been high fire events recorded across the Western States of the USA, especially in California and surrounding regions (Enders et al., 2021). Since the first forest fire incident occurred sometime in the last week of July and the first week of August, August 1 has been considered the starting date for the analysis. *Second*: the active fire events and their geo-coded information were extracted from Soumi-NPP and NOAA-20 fire products from August 1 to October 30. Other auxiliary information such as FRP, brightness temperature, etc., has also been retrieved from the source data. Instead of taking the weekly period, daily active fire events data was utilized for generating FRP hotspots over the space using the kernel density method. *Third:* sensor location information for $PM_{2.5}$, $PM_{10}$, and $NO_2$ was retrieved from the OpenAQ data platform. Stations with no data (or have irregular data such as missing values or very high/low observation) were discarded from the analysis. Using the approach stated above, a total number of 274, 70 and 61 ground air pollution monitoring stations have been identified for $PM_{2.5}$, $PM_{10}$, and $NO_2$. For few monitoring stations, data was found irregular or not available for either year. These stations have also been removed from the final evaluation. The default unit (ppm) of $NO_2$ was converted to $\mu g/m^3$ unit to make it comparable with other pollutants. **Fourth:** The mapping of $PM_{2.5}$, $PM_{10}$, and $NO_2$ was done in two different ways. In the first step, the state-wise average $PM_{2.5}$, $PM_{10}$, and $NO_2$ concentration were measured using the spatial join function in ArcGIS Pro software. The State-wise average concentration of $PM_{2.5}$, $PM_{10}$, and $NO_2$ in the reference year (2019) was provided in supplementary tables. Before putting the pollution estimates into the analysis for spatial mapping and subsequent interpretation, data was checked thoroughly for eliminating outliers and irregular measurements. In the second step, the Inverse Distance Weighted (IDW) kriging was performed using the station's pollution measurements and accordingly, a continuous raster surface was prepared for both fire year and reference year. **Fifth:** The daily concentration of $PM_{2.5}$, $PM_{10}$, and $NO_2$ in the FireStates and ExFireStates from August 1 to October 30 was also incorporated in the assessment to examine the impact of forest fire on air quality of the regions and to assess statistical significant changes in air pollution concentration during the fire period. The distribution and ranges of the air pollution concentration were measured using minimum, first quartile, median, third quartile, and maximum distributions of parameters. Linear trend analysis was also performed for assessing the changes in $PM_{2.5}$, $PM_{10}$, and $NO_2$ concentration during the study period. Linear trend fit and statistical significance of the change estimates were also measured at different probability levels for the defined period for both 2019 and 2020. Daily $PM_{2.5}$, $PM_{10}$, and $NO_2$ values have been used for the linear trend assessment. Daily COVID-19 cases and deaths counts have also been incorporated into the change assessment analysis to examine the association between air pollution concentration and COVID incidents in the fire-prone States of the USA. All the statistical analysis was performed in R statistical software. Multiscale Geographically Weighted Regression (MGWR) was done using the MGWR Python package and MGWRv4 software.

## 3. Results

### 3.1 Forest fire emission and its impact on air pollution

The spatial-temporal changes and variation of different air pollutants, i.e. $PM_{2.5}$, $PM_{10}$, and $NO_2$, were measured using in-situ monitored air pollution data collected from OpenAQ controlled stations. **Fig S1** shows the spatial location and distribution of $PM_{2.5}$, $PM_{10}$, and $NO_2$ ground monitoring stations. Using the spatially distributed monitoring data, the raster map of different air pollutants have been prepared (**Fig. 2, Fig. 3, Fig. 4, Fig. S2, Fig. S3**). A drastic change and upsurges of air pollution concentration ($\mu g/m^3$) are evident in the Western part of the USA that mainly caused by the 2020 West Coast forest fire events. The maximum $PM_{2.5}$ ($\mu g/m^3$) concentration is recorded as ~80, ~250, and ~160 in August, September, and October in 2020, while the same was recorded much lower in concentration, i.e. ~25, ~65, ~45 during the same period in 2019 (**Fig. 2**). A similar changing pattern is observed for $PM_{10}$ and $NO_2$ concentration during the study period (**Fig. 3 and Fig. 4**). $PM_{10}$ concentration ($\mu g/m^3$) is reached up to ~260 in 2020 in the Western states of the USA. In contrast, the concentration of $PM_{10}$ is recorded as much lower in 2019. In addition to the raster based-analysis, the State averaged pollution concentration ($\mu g/m^3$) is also measured for both 2019 and 2020 (**Fig. S2**). In order to save space, only 2020's map has been given in this paper. The state-averaged concertation values of different air pollutants are found comparably higher in the Western states than the rest of the states of the USA. Moreover, the state-averaged change statistics also suggesting a noticeable percentage increases in $PM_{2.5}$, $PM_{10}$, and $NO_2$ concentration during the study period. Among the pollutants, the changes are measured highest for $PM_{2.5}$, followed by $PM_{10}$, and $NO_2$, respectively (**Fig. S3 and Fig. S4**). Additionally, a high percentage of increases are observed in the fire-affected States, i.e. California, Colorado, Washington, Nebraska, etc. (**Fig. S4**). While the other States have shown a negligible to negative (air quality index increases in 2020 due to full/partial COVID lockdown in most of the States) changes in air pollution concentration ($\mu g/m^3$) during August 1 to October 31 period in non-fire (2019) and fire year (2020).

The temporal changes in air pollution concentration during the August 1 to October 30 period is evaluated for both 2019 and 2020 and presented in **Fig. 5, Fig. 6, Fig. 7, Fig. 8, Table. 1, Table. S1, Table. S2, Table. S3, Table. S4, Table. S5, Table. S6**. **Fig. 5** and **Table. S1** shows the monthly averaged concentration ($\mu g/m^3$) of $PM_{2.5}$, $PM_{10}$, and $NO_2$ in August, September, and October months. Only fire states (California, Oregon, Washington, and Colorado) were considered for calculating the monthly statistical distribution of air pollutants in 2019 and 2020. Monthly averaged $PM_{2.5}$ concentration ($\mu g/m^3$) in 2019 was measured as 7.41, 6.79, 5.49, and 5.96 for California, Colorado, Oregon, and Washington, while the same was measured as 20.95, 14.50, 6.46, and 5.09 in 2020, respectively (**Table. S1 and Fig. 5**). For $PM_{10}$, the monthly averaged values ($\mu g/m^3$) were increased in the fire affected states (**Table. S1**). However, for $NO_2$, a decreasing monthly averaged concentration ($\mu g/m^3$) was observed during the study period (**Fig. 6** and **Table. S1**). Changes in air pollution concentration due to forest fire are also examined and presented in **Fig. 6, Fig. 7, Table. 1**, and **Table. S2**. During the entire study period (August 1 to October 30), changes in $PM_{2.5}$ was measured highest for Oregon, followed by Washington, California, and Colorado, respectively (**Table. S2**). For $PM_{10}$, the changes (positive) were highest in Washington, while the same measured comparably low in California and Colorado states (**Table. S2**). Among the three air pollutants, changes in percentage concentration were found lowest for $NO_2$, as many States have shown the declining status of $NO_2$ concentration in 2020 (**Table. S2**). This happened due to the partial/fully lockdown imposed in many states due to the outbreak of COVID in 2020. In addition to this, the State-wise monthly averaged concentration of $PM_{2.5}$, $PM_{10}$, and $NO_2$ is also measured and presented in **Fig. 6, Table. S3**, **Table. S4**, **Table. S5**. These tables and Figures collectively suggest that $PM_{2.5}$, $PM_{10}$, and $NO_2$ concentrations were increased in 2020, mainly due to the record-breaking forest fires in the West Coast regions of the contiguous US in 2020. A linear trend curve is also fitted for three air pollutants for both 2019 and 2020 to examine the changes in daily $PM_{2.5}$, $PM_{10}$, and $NO_2$ concentration during the study period. The daily averaged linear fit surface measured for both 2019 and 2020 was found significantly high in 2020 than that of 2019. Additionally, the peak $PM_{2.5}$ and $PM_{10}$ concentration ($\mu g/m^3$) have reached up to ~100 (for $PM_{2.5}$) and 200 (for $PM_{10}$), whereas the daily concentration of the pollutants was measured much lower in 2019, reached maximum up to 25 (for $PM_{2.5}$) and 150 (for $PM_{10}$), respectively (**Fig. 7**). Statistical non-parametric test has been done to examine the mean difference in $PM_{2.5}$, $PM_{10}$, and $NO_2$ concentration between fire (2020) and reference (2019) year and presented in **Table. 1**. The mean differences of all three air pollutants, i.e. $PM_{2.5}$ (reported as model 1, 4, 7), $PM_{10}$ (model 2, 5, 8) and $NO_2$ (model 3, 6, 9) between fire and the non-fire year was found statistically significant at significance level $P<0.05$ (**Table. 1**). In continuation to the comparison analysis between the air pollution concentration in fire and non-fire years, the monthly averaged air quality index (AQI) is also computed using EPA's defined AQI threshold values. AQI has been found to deteriorate in 2020 in California, Colorado, Idaho, Oregon, Wahington, etc. (**Table. S6**). These States have been affected badly by the 2020 forest fire, and a considerable amount of forest area (acres) has been lost that resulted in adding a substantial amount of fine and coarse particulate matter into the atmosphere. The added concentration of $PM_{2.5}$, $PM_{10}$, and $NO_2$ that has been observed in this study could be entirely due to forest fire emission, as most the States and large cities in the contiguous USA undergone strict lockdown in 2020 to prevent the spread of COVID-19, which means the contribution of anthropogenic emission that comes from traffic and industry is halted and therefore the same has negligible contribution to the increases of air pollution concentration in 2020.

To analyze the national scale air pollution status in the entire USA during 2000 to 2019 and to compare the trend of changes during 2000 – 2019 period with 2020 forest fire emission status, the linear trend plots have been drawn for both the key air pollutants ($PM_{2.5}$, $PM_{10}$, and $NO_2$) (**Fig. S5**), as well as for the fire determinant climatic variables, i.e. average precipitation, maximum and minimum temperature, etc. (**Fig. S6**). **Fig. S5** shows the linear temporal changes in $PM_{2.5}$, $PM_{10}$, and $NO_2$ concentration in the entire USA (**Fig. S6a**), Western region (**Fig. S6b**), and South Western region (**Fig. S6c**), respectively. $PM_{2.5}$ concentration has been reduced significantly during the 2000 – 2019 period for all three regions. However, since the Western part of the USA have frequently received large scale forest fires periodically, the changes in $PM_{2.5}$ concentration during 2000 - 2019 was found much lower in the Western region ($R^2 = 0.72$), while the same was measured comparably higher at the National level ($R^2 = 0.96$) and South Western region ($R^2 = 0.86$), respectively (**Fig. S5**). For $PM_{10}$, the changes were measured much higher ($R^2 = 0.71$) at the national level, calculated as 104 ($\mu g/m^3$) in 2000 and 56 ($\mu g/m^3$) in 2019 than that of the Western region ($R^2 = 0.63$). However, simultaneously, the changes in $PM_{10}$ concentration during 2000 – 2019 were found insignificant in the South Western region (**Fig. S5**). Additionally, $NO_2$ concentration during 2000 – 2019 has been reduced significantly, found maximum changes for the national level ($R^2 = 0.97$), followed by Western ($R^2 = 0.95$), and South Western region (0.91), respectively (**Fig. S5**). The increasing pattern of both mean and maximum temperature might have augmented the impact of forest fire in the West Coast region of the USA (**Fig.**

S6). Loess filter has been used to smoothen the data collected for the period 1895 – 2020 for all three climatic variables. In addition to the temperature variables, average precipitation has also been increased during the 1895 – 2020 period (**Fig. S6, Table. S7**). Apart from the key air pollutants, the emission of other pollutants was estimated using the GFED. V4.1 database (**Table. S8, S9**). Among the major biomes, only all fires and temperature biome fires were considered for effective comparison analysis. Due to the massive forest fire events in 2020, the emission of Carbon, Black Carbon, Carbon Monoxide, Carbon Dioxide, Nitrous Oxide, Nitrogen Oxide, Particulate Matter, Sulphur dioxide, Total Particulate Matter, and Carbon from total particulate matter was increased substantially, and the measured values were found much higher in 2020 compared to the 1997 – 2019 period (**Table. S8**). The year-wise summary values of different air pollutants were also measured and presented in **Table. S9**. Emission of different air pollutants was measured comparably very high in 2020 than the preceding years (**Table. S9**). The number of large forest fire events and associated emission of Carbon and fine particulate matters could be the main reason for these exceptionally high emission estimates that were evident in the Western part of the USA in mid to late 2020.

## 3.2 Forest fire led emission and its association with COVID counts

**Fig. 8** shows how COVID new cases/death counts and the concentration of $PM_{2.5}$, $PM_{10}$, and $NO_2$ have changed on a daily scale during the study period. For California and Colorado, a synergistic association between the daily $PM_{2.5}$, $PM_{10}$, and $NO_2$ concentration and COVID numbers are observed. For the remaining two States, i.e. Oregon and Washington, no such close association between the fire emission and COVID counts have observed (**Fig. 8**). To further extend the correlation analysis, a correlation matrix has been drawn consisting of month-wise distribution of air pollution estimates and COVID counts in the fire States (**Fig. 9, Fig. 10**). **Fig. 9** was illustrated using the COVID and air pollution values individually for each of the four fire States, and **Fig. 10** was plotted based on the averaged values of COVID counts and $PM_{2.5}$, $PM_{10}$, and $NO_2$ concentration of the four fire States. COVID cases and death were statistically significantly correlated with $PM_{2.5}$ October, $PM_{10}$ October, $NO_2$ September and NO2 October estimates (**Fig. 9**). However, considering the average COVID-19 numbers and air pollution values of the four fire States, a moderate association between the COVID counts and air pollution was found for all three months considered in this study (**Fig. 10, Table. 2**).

The outcomes of the spatial regression analysis between the averaged (average of three months, i.e. August, September, October) air pollution values and COVID-19 counts are presented in **Fig. 11** and **Fig. 12**. For all the test case experiments between COVID counts and air pollution estimates, comparably high spatial $R^2$ values are measured for the State of Colorado and California counties. In contrast, lower spatial $R^2$ values are measured for the counties in Washington and Oregon (**Fig. 11 and Table. 3**). This implies a spatial non-stationary and localized association between the explanatory (air pollution in the present study) and response variables (COVID case and deaths) which can not be explained through a global stationary regression model. A similar pattern of association was found between the county averaged maximum $PM_{2.5}$, $PM_{10}$, and $NO_2$ concentrations and COVID-19 counts (**Fig. 12 and Table. 4**). For the month-wise analysis, a total of 18 spatial models were developed that covered both COVID cases and death and accompanied with relevant model diagnostics tests, including AIC, BIC, adjusted t-test, to assess the statistical significance of the models at different probability level and uncertainty estimates that are associated with different model parametrization (**Table. 3**). To capture the overall local association between the explanatory and response variables for the entire study period (August 1 to October 30), a total of 12 spatial regression models were developed by considering the averaged values of the COVID-19 and air pollution estimates (both averaged and maximum values were considered) of the studied fire States (**Table. 4**). Among the models, the highest local $R^2$ are measured for NO$_2$max ($R^2 = 0.542$ for cases and $R^2 = 0.556$ for death) and NO$_2$mean ($R^2 = 0.409$ for cases and $R^2 = 0.443$ for death) (**Table. 4**). In addition to the spatial regression, the negative binomial regression is also performed to examine the effect of dispersion into the modelling outcomes (**Table. S10** for COVID-19 case and **Table. S11** for COVID-19 death). A total of two NB models were performed to examine the association between the explanatory and response variable with adjusted dispersion effects on the model. For both the models, the α or the estimate of dispersion parameter are found greater than 0, which implies the presence of overdispersion in the data. The positive coefficient values suggest that one unit increases of the predictor variable lead to x unit change in the expected outcome of the response variable. Among the model combinations, the positive coefficients values were found between the PM$_{2.5}$max, PM$_{10}$mean, and NO$_2$max with COVID19 cases **Table. S10, S11**. These suggest that one unit changes in PM$_{2.5}$max, PM$_{10}$mean, and NO$_2$max would increase the COVID-19 cases by 0.014, 0.078, and 0.251 units, respectively. While for the death factor, the coefficient values are measured as 0.085 and 0.260 for PM$_{10}$mean and NO$_2$max (**Table. S10, S11**).

**Discussion**

The drastic increases of air pollutants in the mid to late 2020 in the West Coast States of the US indicates that the fires in 2020 not only caused severe impact on wildlife and structural damages of properties, but they also added high amounts of gaseous and particulate pollutants including smoke and ash into the atmosphere, raising health emergency amidst the COVID-19 pandemic period. Though the clear connection between health effects and forest fire smoke has not fully explored, there are substantial evidence that supports the strong association between fire emitted smoke and ash concentration and severe health outcomes. Research has also observed that such effects have often exhibited delayed/lagged consequences, especially for cardiovascular and respiratory cause-specific diseases, and do not usually disappear when air gets clear (Landguth et al., 2020). Due to forest fire and the associated emission of particulate matter, a positive trend in $PM_{2.5}$ and $PM_{10}$ concentration was observed in the North West and Western United States (the most fire-prone region in the USA), compared to the other areas of the country for which a negative trend in $PM_{2.5}$ was recorded during the study period. This positive trend in $PM_{2.5}$ concentration was mainly associated with forest fire, black carbon emission and emission of smoke in the West Coast region. The forest fire smoke emission above the boundary layer is often elevated into the free troposphere through convective lofting where the highspeed winds help travel the smoke to a long distance. The subsidence of this smoke layer occurs at the zone of high surface pressure, which can contribute to the enhancement of surface concentration of fine particulate matter (Miller et al., 2011).

There are many factors responsible for recurring forest fire events in the USA's West Coast region. The current year has witnessed the above-normal heatwave (~54 °C temperature recorded in some places in California in 2020), which triggered and created the ideal condition for ignition's, which eventually led to massive forest fires. On the other hand, the cold breezes from neighbouring states, Colorado, have supplied the winds that helped grow the blazes across the region. In line with the previous studies, the present analyses found a comparatively higher forest fire activity, mostly over Western regions of the US, followed by a significantly uplifting PM concentration. At the same time, other parts of the country exhibited a substantially lower concentration of atmospheric pollutants due to the strict practice of lockdown. Additionally, despite being located in the Coastal region where local wind breeze often subdues the high pollution concentration, the Western Coastal states of the USA exhibited high spikes of particulate matter and $NO_2$ concentrations amidst the COVID lockdown period when the anthropogenic emission is partially or entirely switched-off.

The unprecedented and abrupt increase of fine and coarse particulate matter in 2020 has mainly happened due to the forest fire emission. This added concentration in $PM_{2.5}$ and $PM_{10}$ could pose serious health concerns to the residents living in this region. Forest fire smoke causes 339,000 annual global premature mortality (interquartile range: 260,000–600,000) (Johnston et al., 2012). An earlier study that analyzed 61 epidemiological studies linked with a forest fire and human health across the world and reported that daily air pollution levels recorded during or after forest fire events exceeded US EPA regulations (Liu et al., 2015). In many cases, the average $PM_{10}$ concentration during the forest fire period was found 1.2 to 10 times higher than that of non-fire periods (Liu et al., 2015). Among the diseases primarily attributed to forest fire, the respiratory disease was highly associated with forest fire smoke concentration (Liu et al., 2015). In the USA, nearly 10% of the population (30.5 million) stayed in the regions where the contribution of forest fire to annual average ambient $PM_{2.5}$ was high (>1.5 μg/m$^3$). Nearly 10.3 million people experienced unhealthy air quality for more than 10 consecutive days attributed to forest fire and resulted in emissions (Rappold et al., 2017). People with the existing respiratory illness have found to be more susceptible to forest fire air pollution effects, and the upsurges of the demand for rescue medication have been linked to the average exposure time to forest fire smoke in Southern California, USA (Vora et al., 2011). Ischemic Heart Disease (IHD) has also been linked to forest fire emission (particularly $PM_{10}$ emission) as an elevated number of IHD related clinic visits were reported during the forest fire seasons in California (Lee et al., 2009).

The statistically significant association between the fire emitted particulate matter and COVID numbers measured from both spatial and non-spatial regression models indicated the strong impact of forest fire and resulted in emission of air pollutants on human health. Since the West Coast forest fire has begun in the mid to late 2020, this gives us a once in a century opportunity to carefully examines the close connection between the two extreme events, i.e. record-breaking forest fire in 2020 and the associated emission of air pollutants and its effects on COVID-19 casualties. Since several studies have already established the fact that high concentration of air pollutants, especially $PM_{2.5}$, $PM_{10}$, and $NO_2$, can exacerbate the spread and overall casualties of COVID-19 across the scale, it is expected that the added amount of particulate matter and $NO_2$ concentration will have a significant impact on the overall COVID scenarios of the country. A table has been prepared that summarizes previous research findings and shows how different air pollutants have affected COVID spread and overall casualties caused by COVID-19. Among the studies, Wu et al. (2020) analyzed COVID data from 3,000 counties in the USA and found that the increases in 1 $\mu g/m^3$ $PM_{2.5}$ were associated with increased COVID case by about 15% (**Table. 5**). The results of the negative binomial regression of the present study have also indicated a similar association between the predictors and response variables. This study found that both average and maximum concentration of the key air pollutants are highly associated with the COVID cases and deaths in the fire States considered in this study. More detailed analysis with updated COVID-19 data and air pollution measurements will allow us to explore the actual and lagged impact of the 2020 forest fires on both the atmosphere and COVID casualties.

**Conclusion**

The present research has evaluated the effects of forest fire on air quality and COVID-19 casualties in the West Coast states (California, Colorado, Oregon, and Washington) in the USA. In 2020, the West Coast forest fires broke the past forest fire records, and nearly 8 million acres forest areas were burned in California, Colorado, Oregon, and Washington states in the USA. To understand the adverse effects of forest fire emissions on air quality, the concentrations of different air pollutants, i.e. $PM_{2.5}$, $PM_{10}$, and $NO_2$, were measured using in-situ monitored data for a fixed period (August 1 to October 30) for 2020 (fire year) and 2019 (reference year). Additionally, both spatial and non-spatial regression analysis is performed using the county scale COVID-19 and air pollution data to better understand the adverse effects of fire emission on human health. The concentration of particulate matter and $NO_2$ increased from three to five times during the fire period in 2020. The abundance of dry fuel load and increasing surface and air temperature has portrayed the potential risk of large scale fire events in the Western region of the USA. Due to the unprecedented forest fire and particulate matter, and gaseous emissions, the average level of exposure to hazardous pollutants has been increased significantly. This may cause severe health hazards if the fire smoke persists over an extended period. The high-intensity fire events in West Coast regions have been triggered by many causal factors, including heatwaves and a warming climate, the abundant load of dry fuel and lack of soil moisture, cold breezes from nearby states and the occurrence of hurricanes amid the fire events. The increased level of air pollution caused by the West Coast forest fire in 2020 revealed a statistically significant positive association with the COVID-19 casualties. This suggest that forest fire and resulted in gaseous and particulate matter emission could play a major role if the same occurs together with an pandemic or epidemic situation.

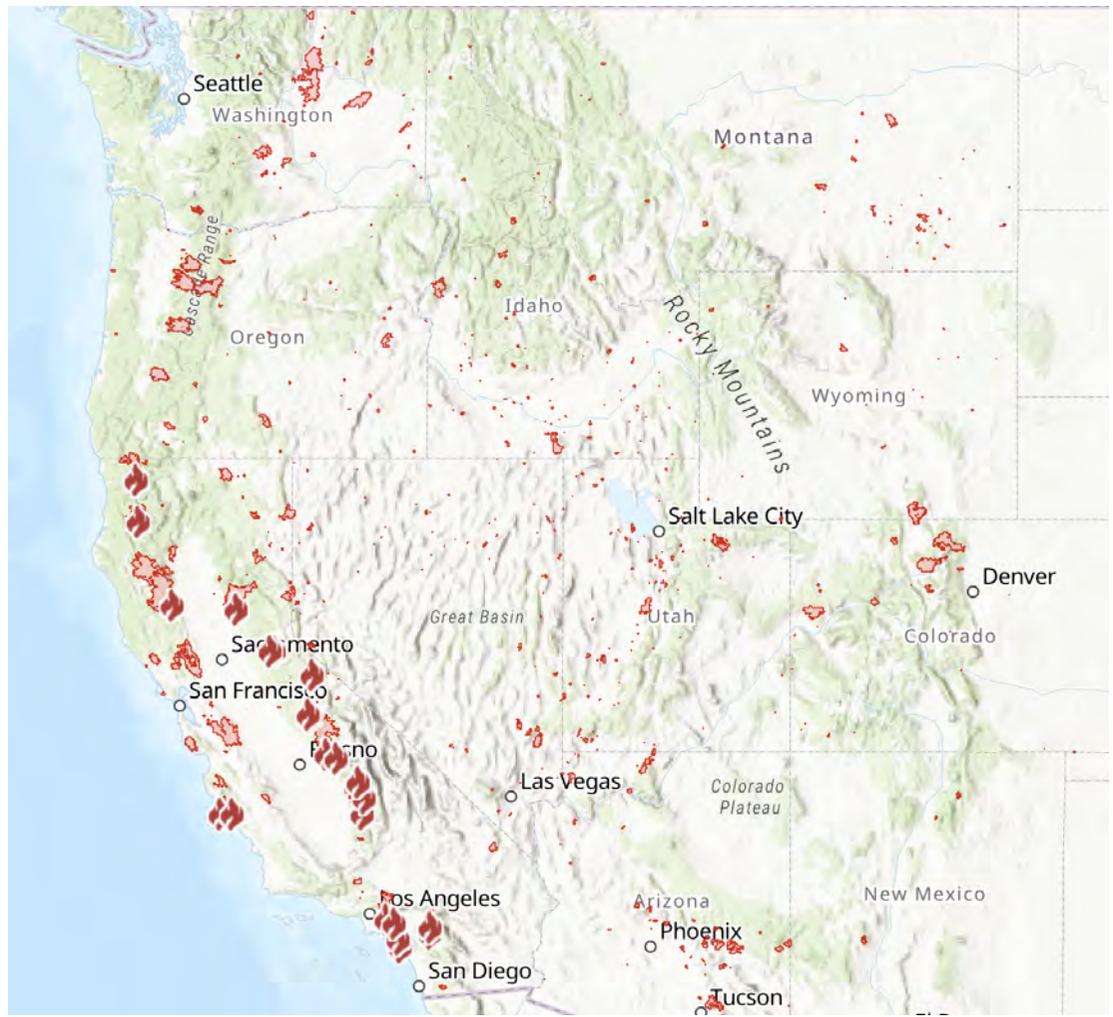

**Fig. 1** Major fire events in California, Oregon, Washington, and Colorado

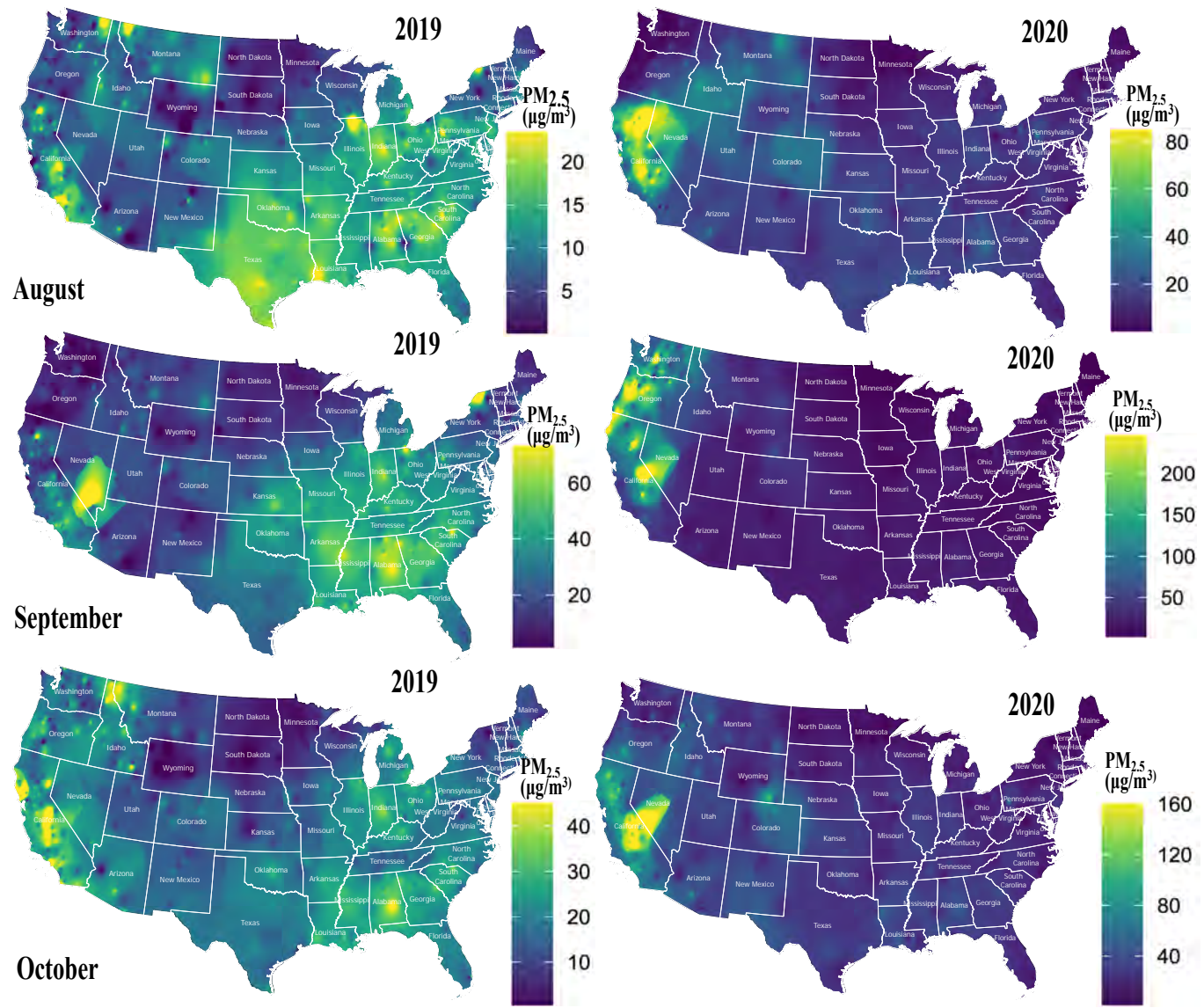

**Fig. 2** Spatial distribution of PM$_{2.5}$ concentration in August, September, and October in 2019 and 2020.

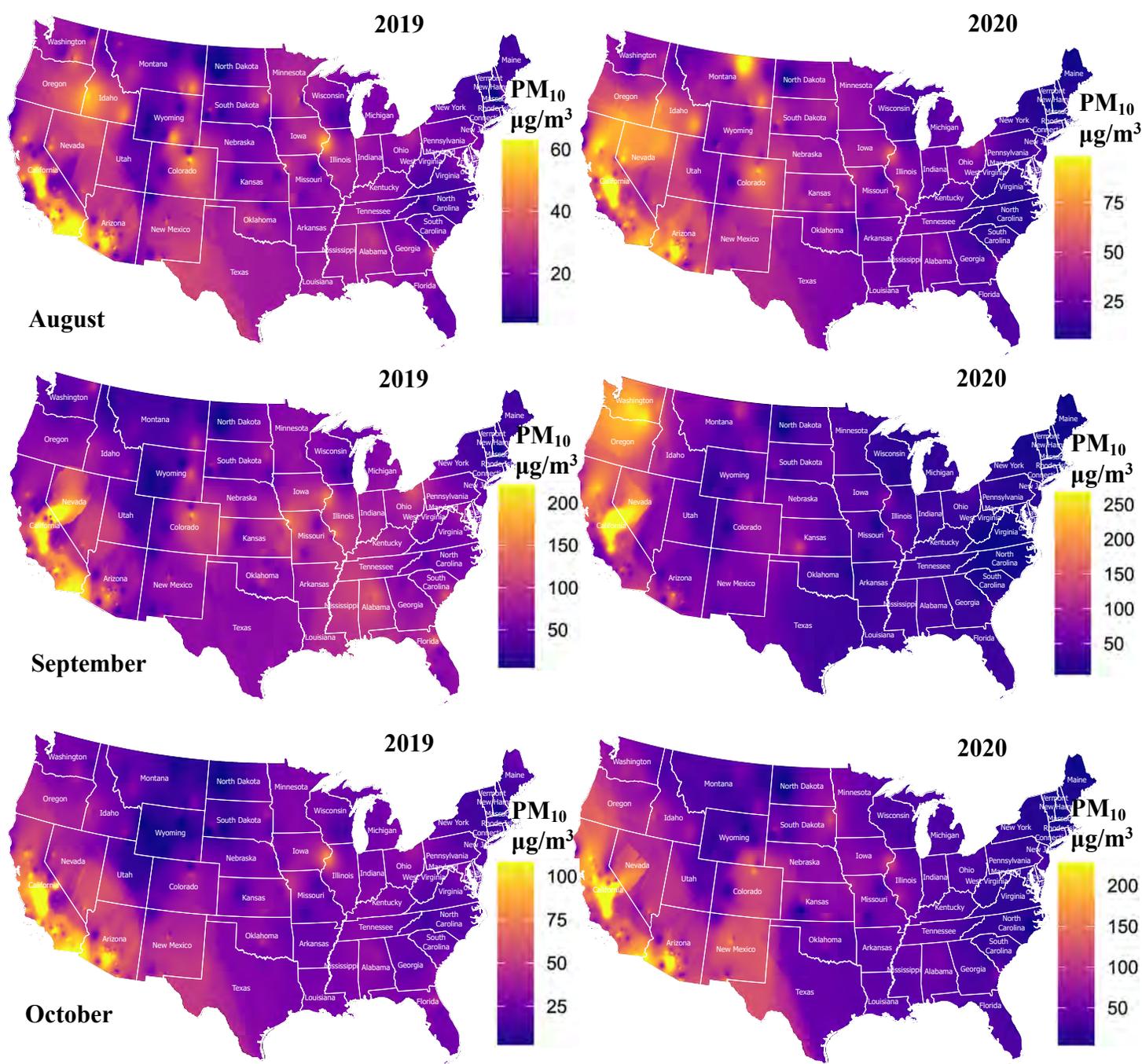

**Fig. 3** Spatial distribution of PM$_{10}$ concentration in August, September, and October in 2019 and 2020.

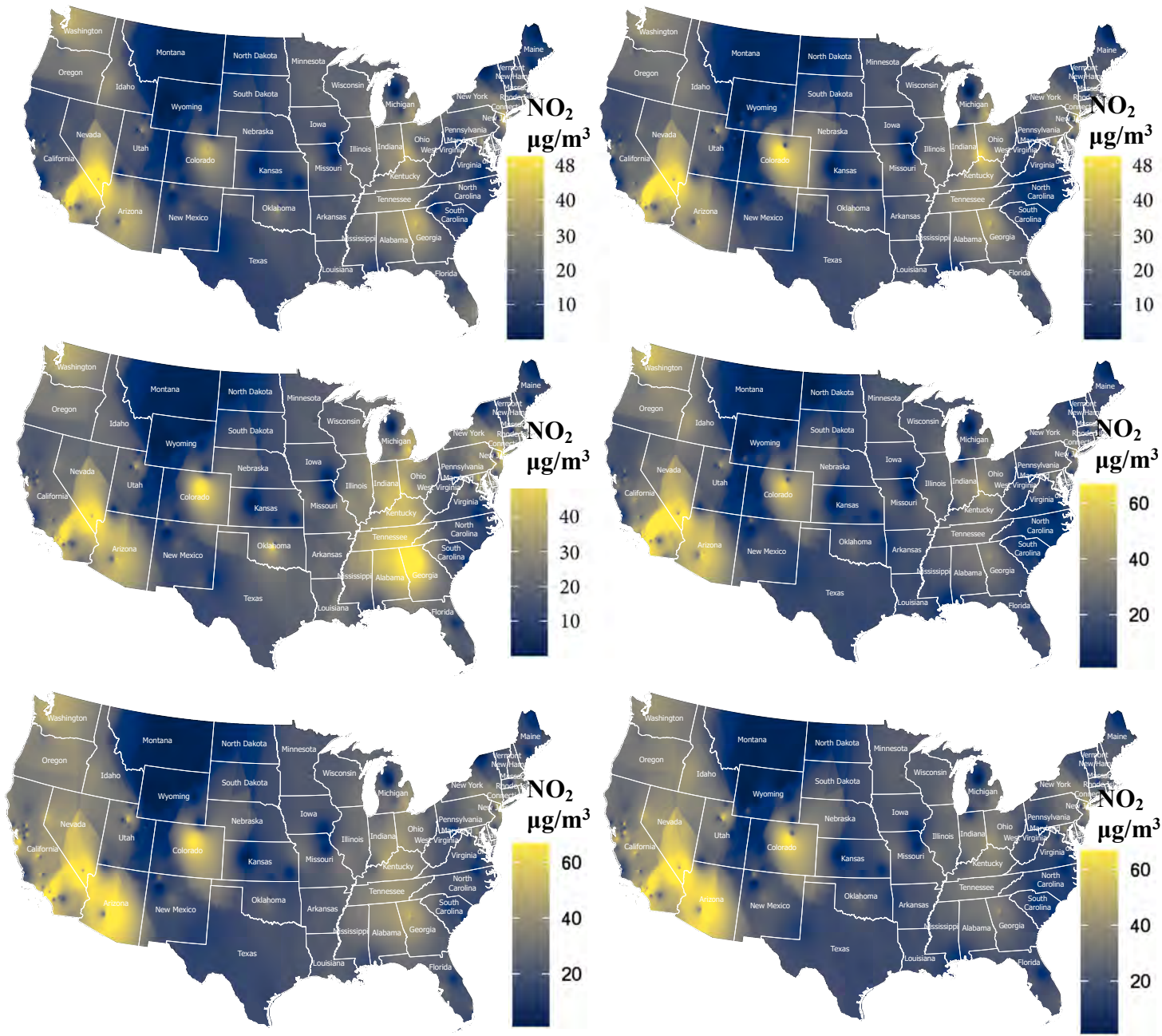

**Fig. 4** Spatial distribution of NO2 concentration in August, September, and October in 2019 and 2020.

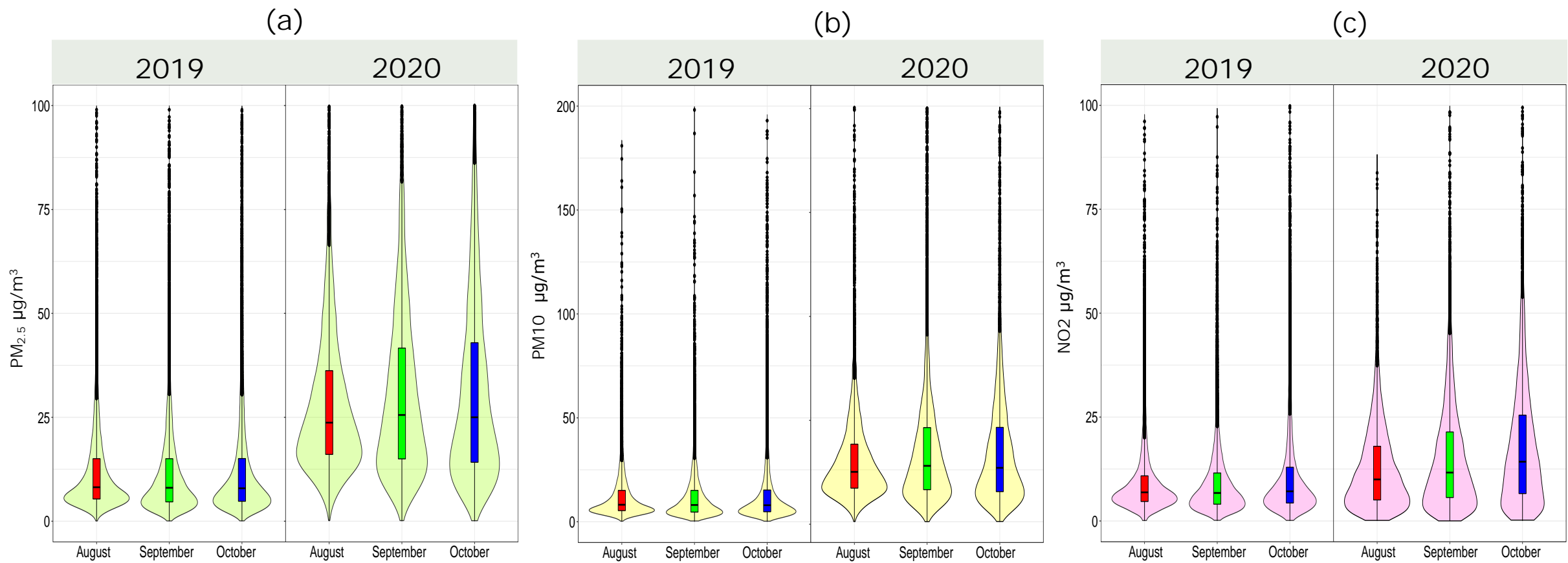

**Fig. 5** Average concentration of different air pollutants (PM2.5, PM10, NO2) in August, September, and October in 2019 and 2020.

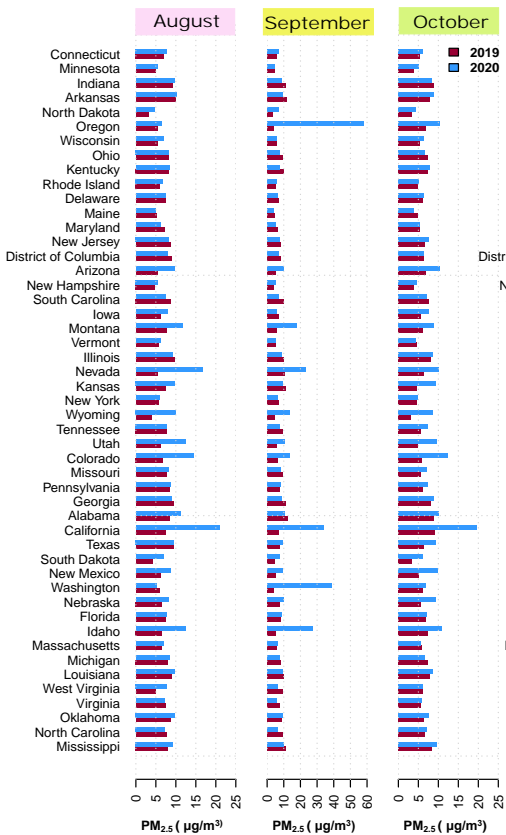
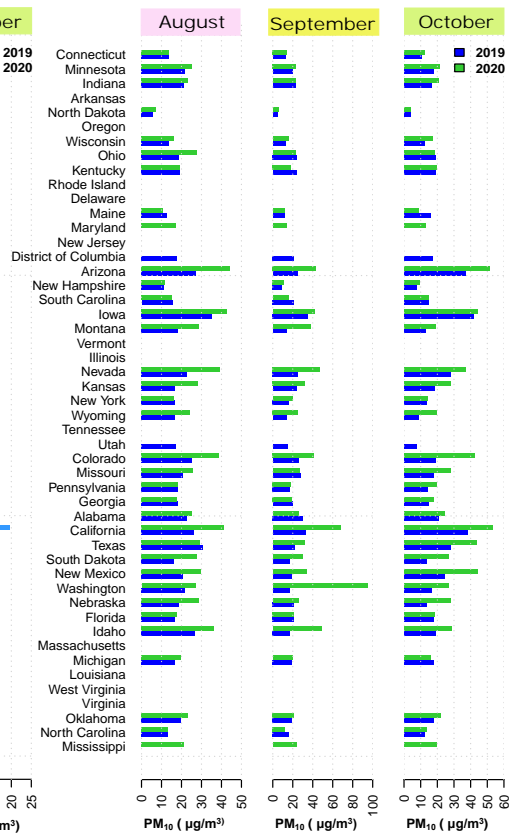
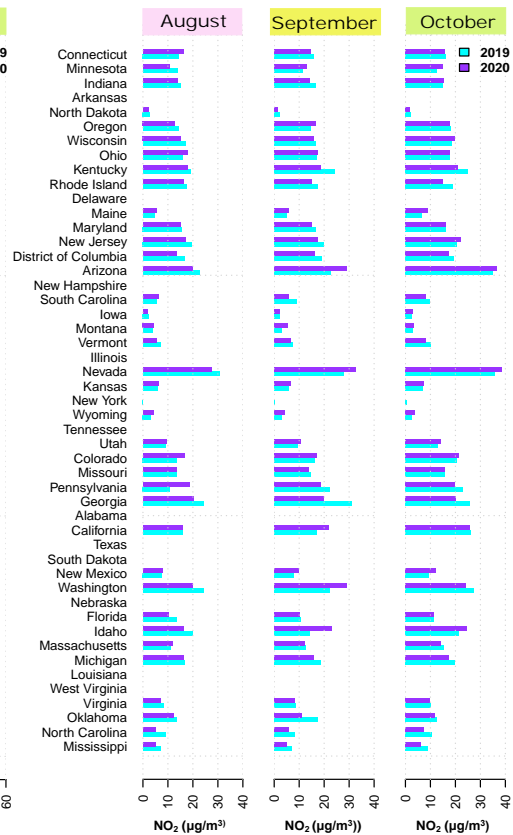

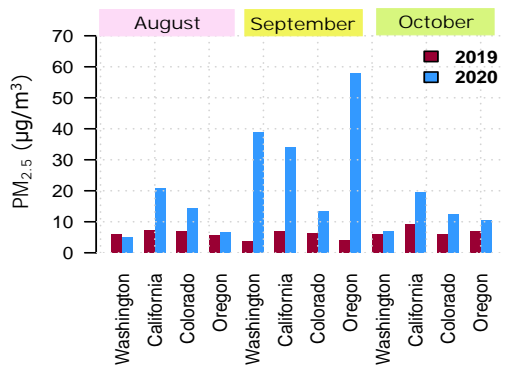
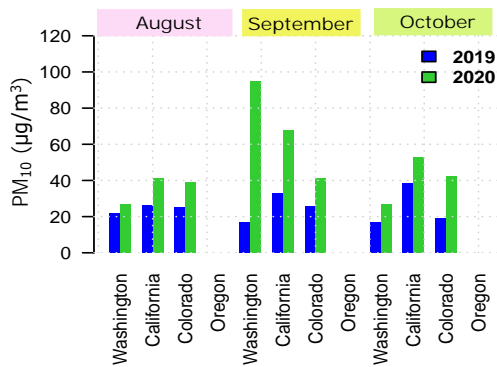
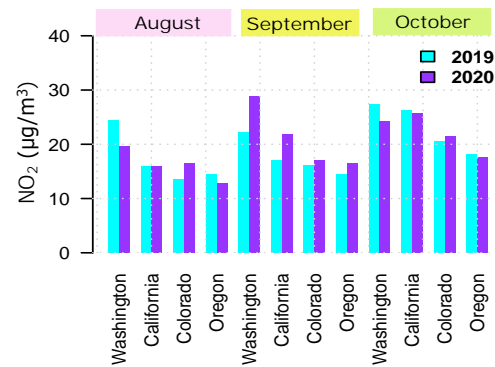

**Fig. 6** Average concentration of (a) PM$_{2.5}$, (b) PM$_{10}$, (c) NO$_2$ in different States in USA in 2019 and 2020; and concentration of (d) PM$_{2.5}$, (e) PM$_{10}$, and (f) NO$_2$ in fire prone States in the USA in 2019 and 2020.

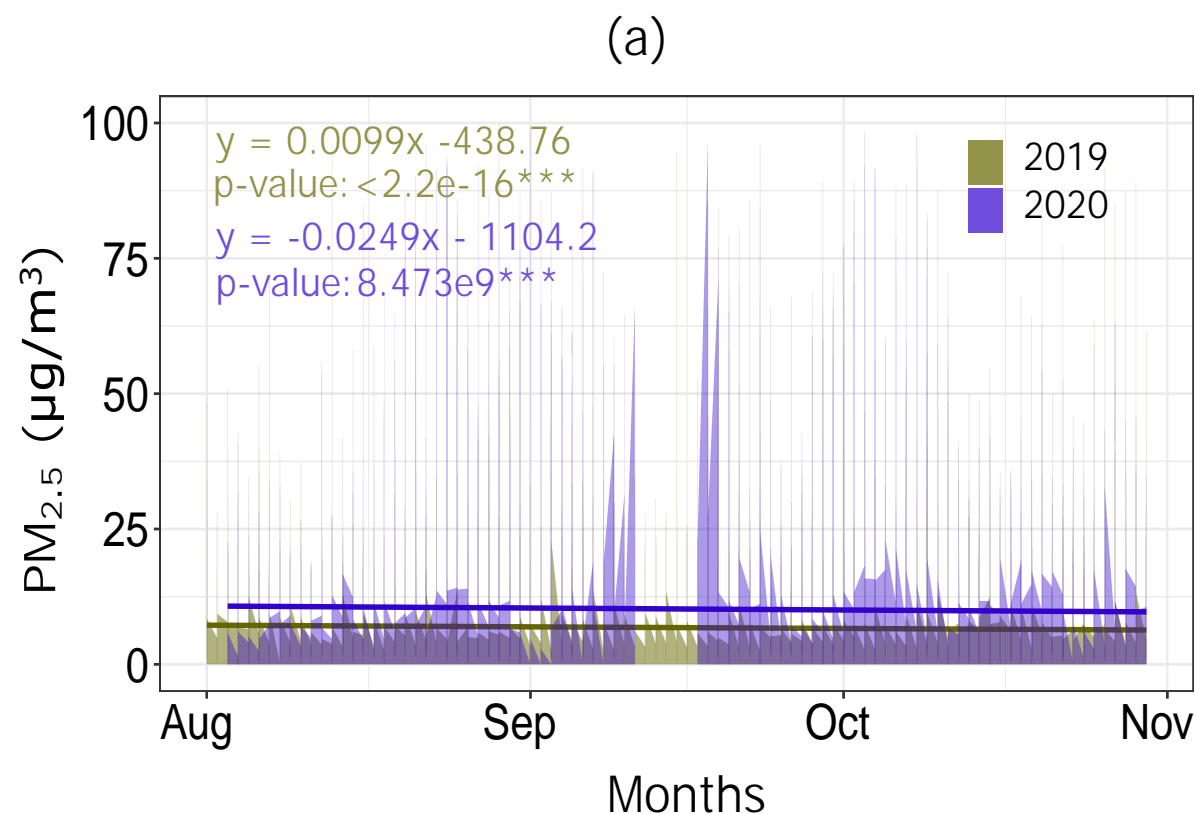
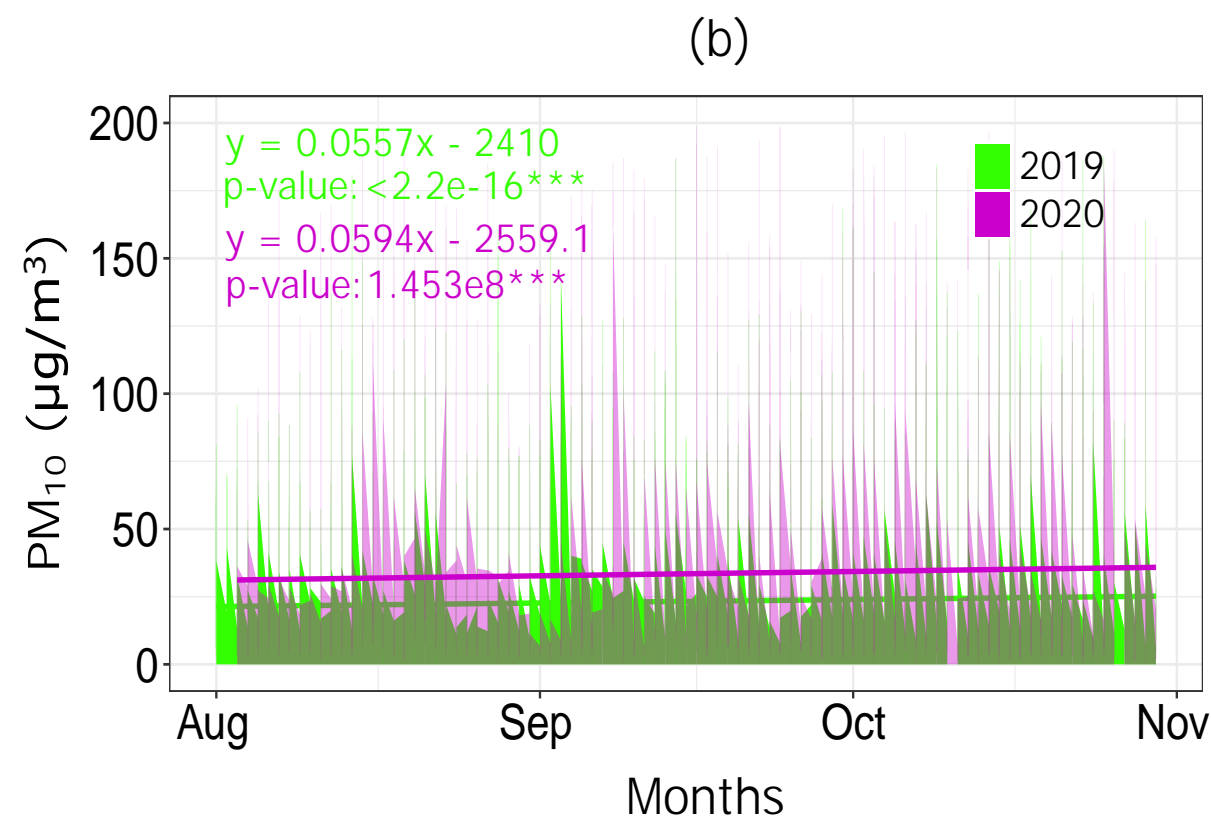
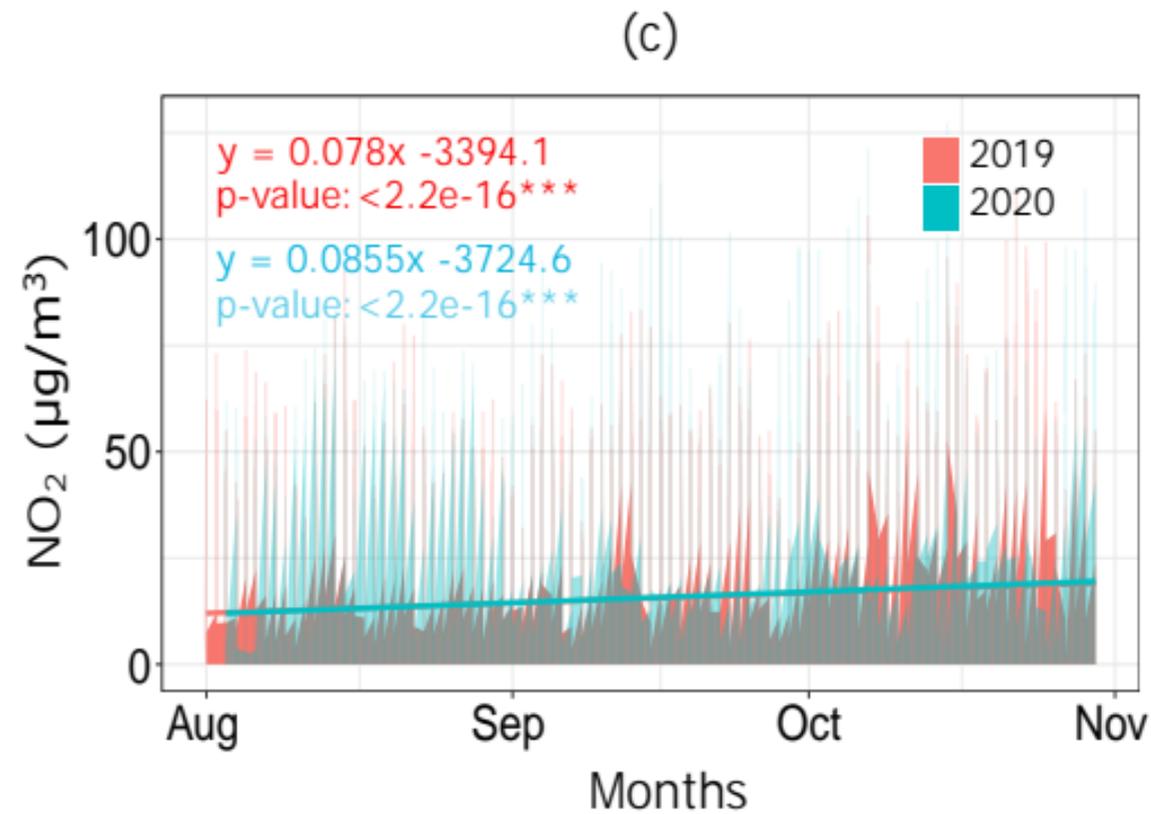

**Fig. 7** Daily variation in PM$_{2.5}$, PM$_{10}$, and NO$_2$ concentration during August 01 to October 30 period in 2019 and 2020.

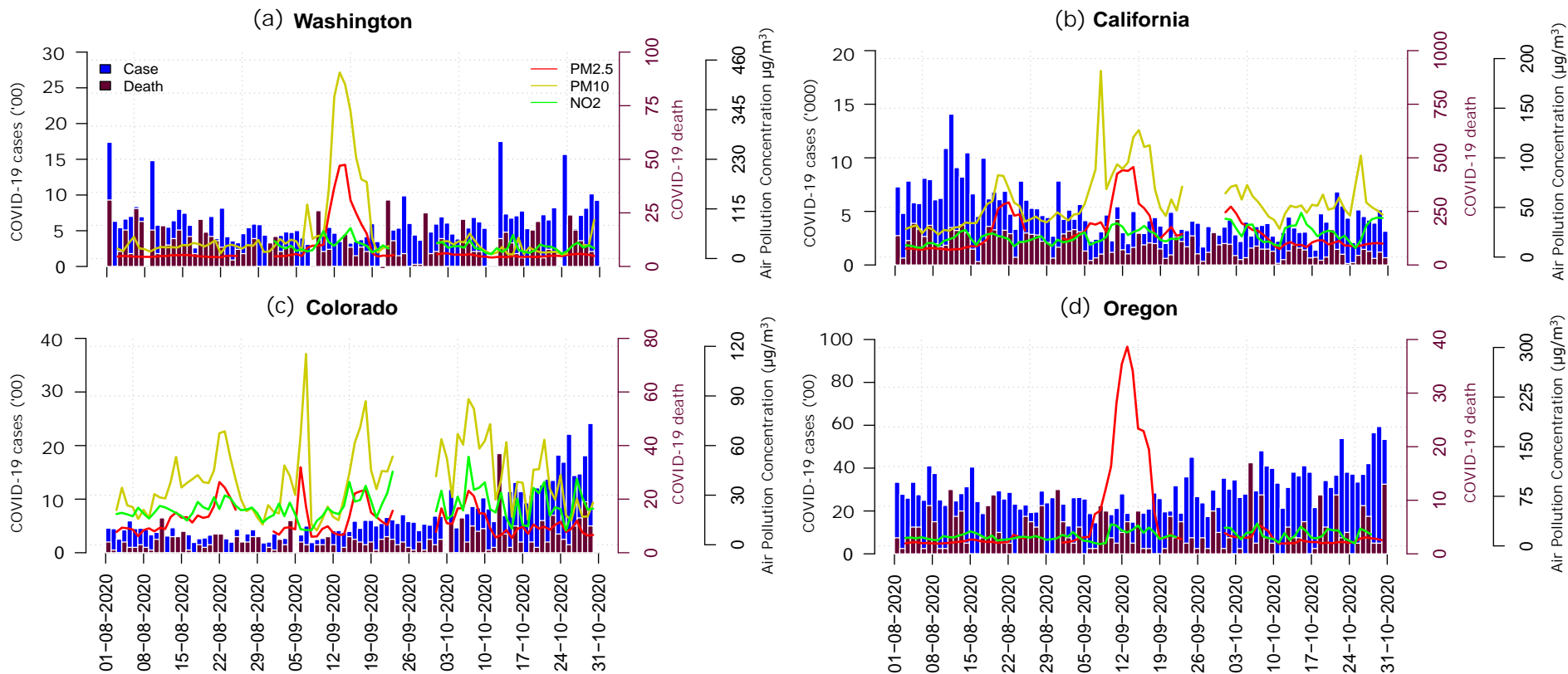

**Fig. 8** Daily changes in air pollutants (PM$_{2.5}$, PM$_{10}$, NO$_2$) and COVID-19 new cases and deaths in the four fire States during August 1 to October 30 period.

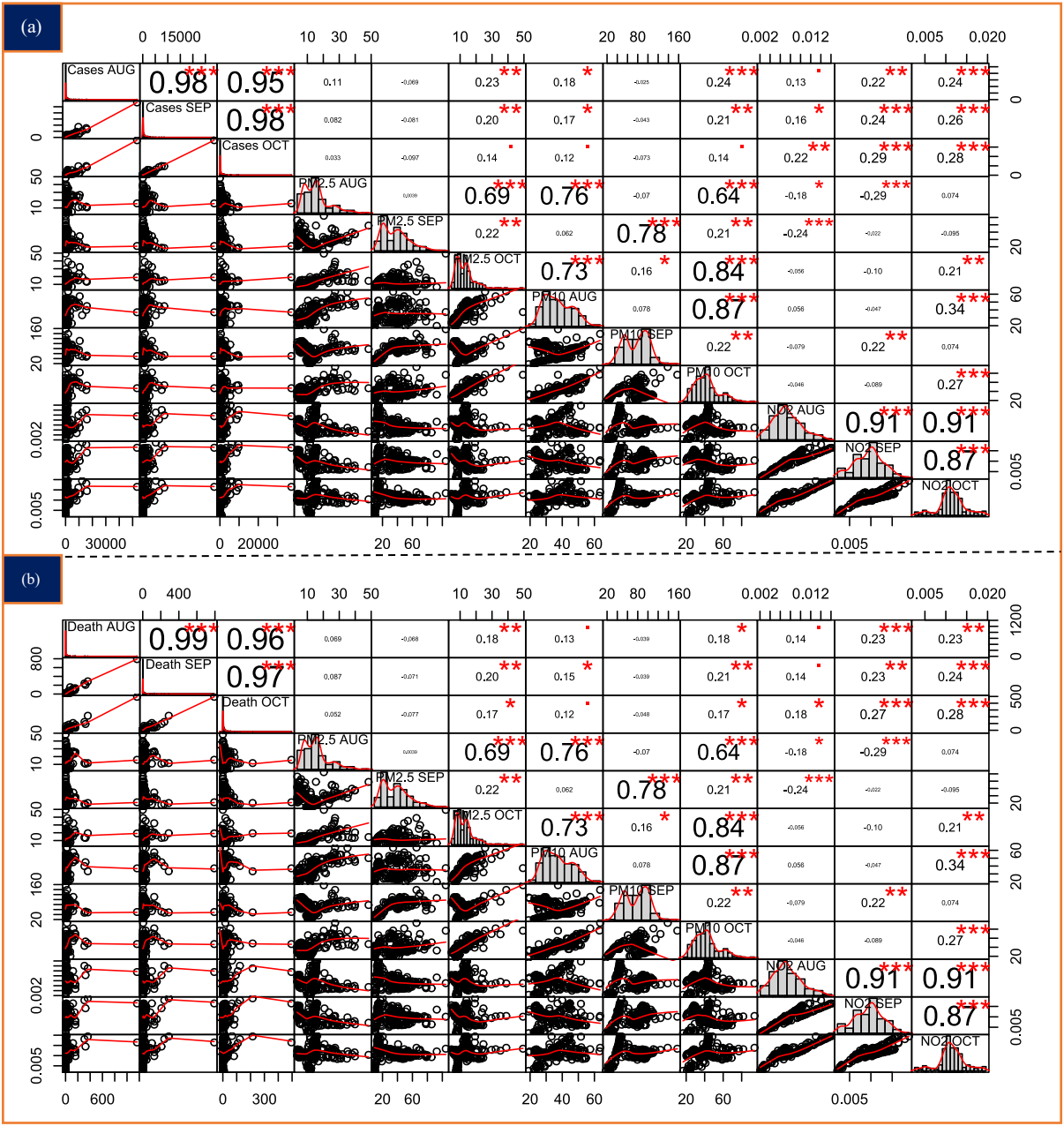

**Fig. 9** Correlation matrix shows the association between monthly concentration values of PM2.5, PM10, and NO2 and monthly average COVID incidents.

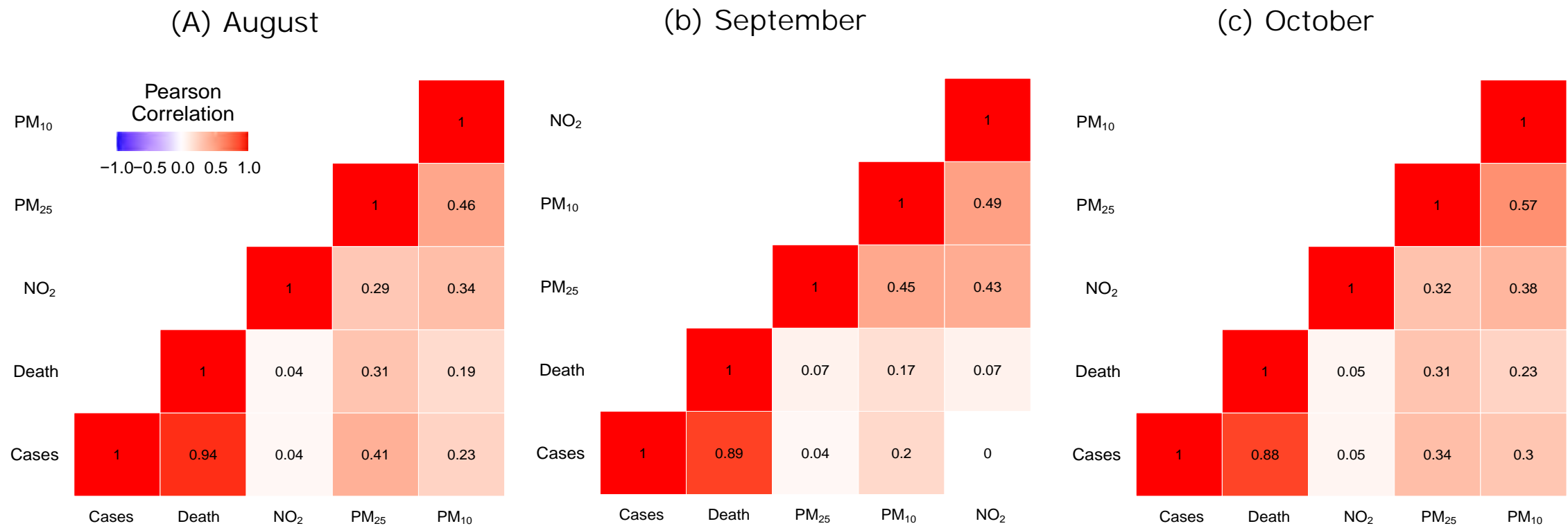

**Fig. 10** Correlation matrix shows the association between the air pollution estimates and COVID-19 casualties duing the entire study period (August 1 to October 30).

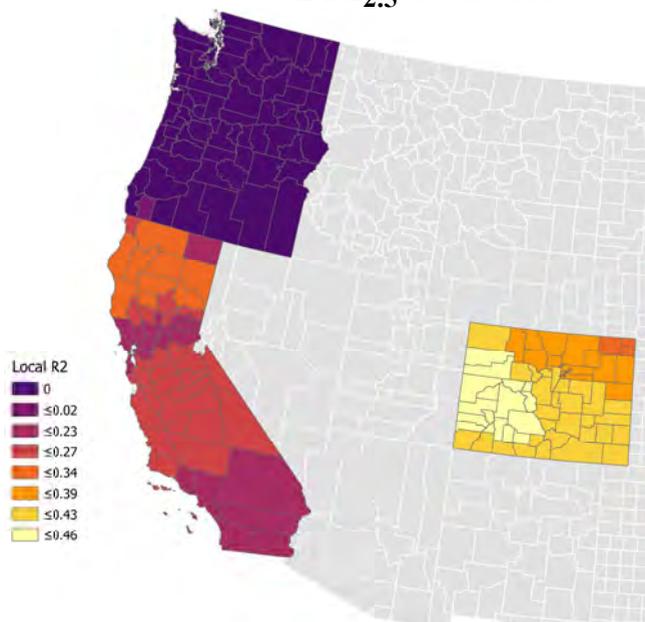

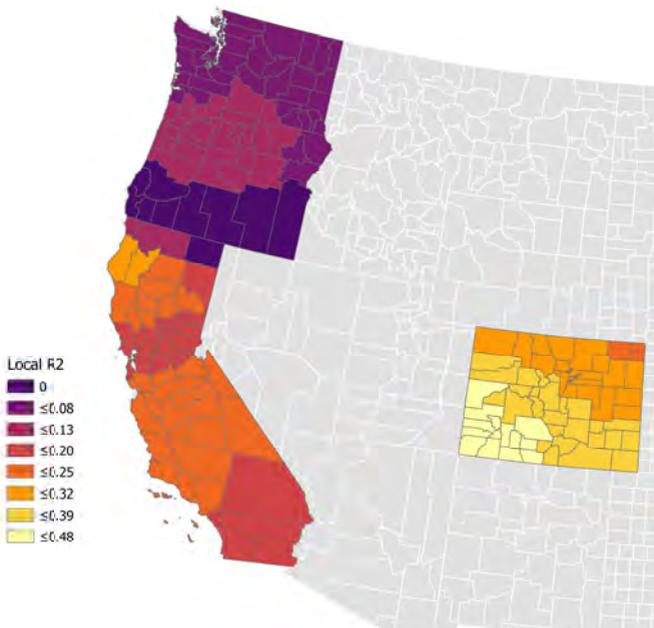

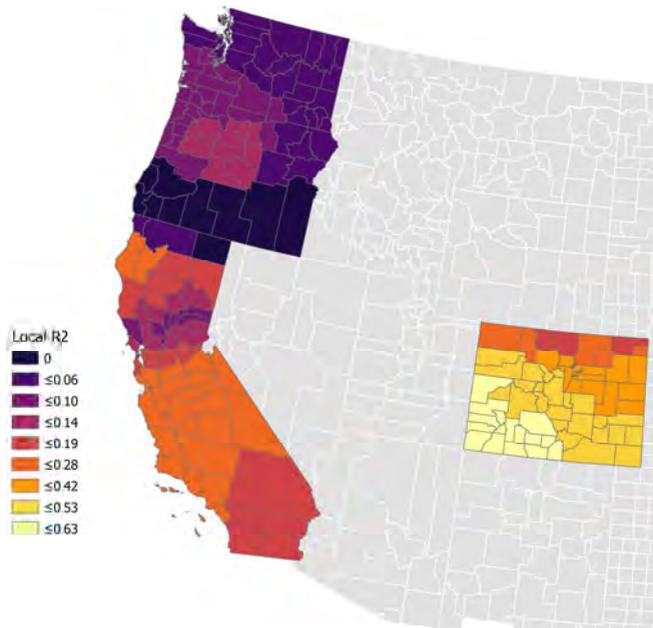

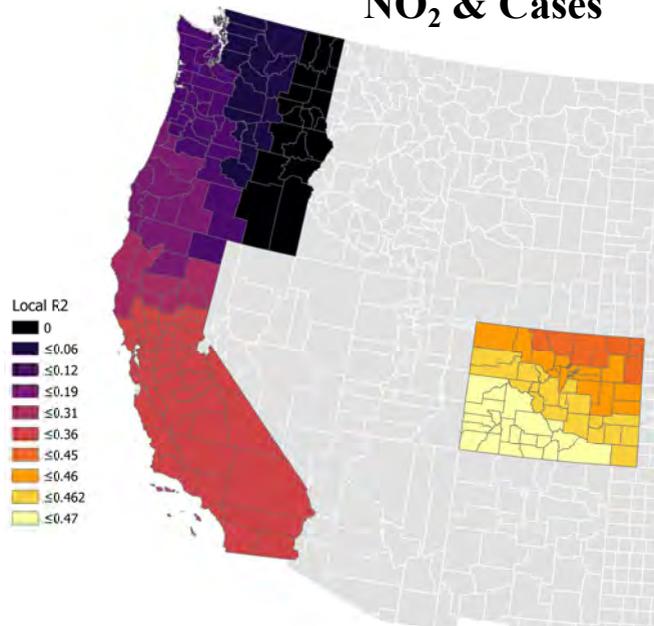

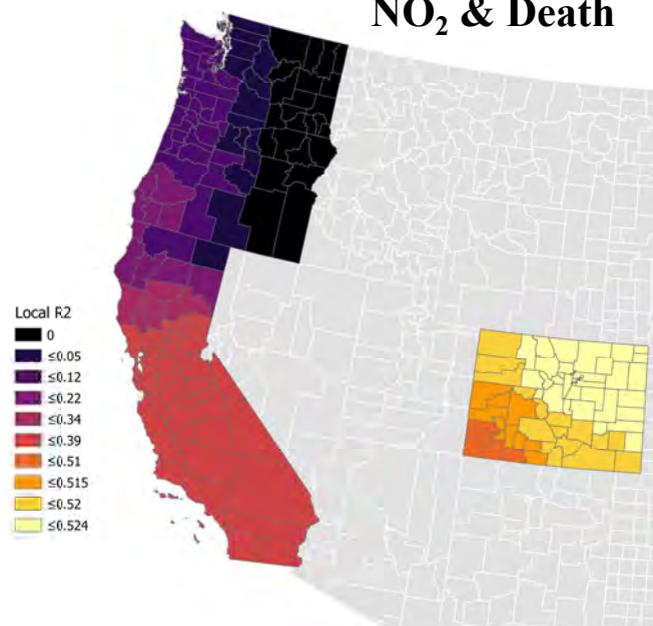

**Fig. 11** Spatial coefficient of determination values ($R^2$) exhibiting the spatial association between air pollution estimates and COVID incidents at local scale.

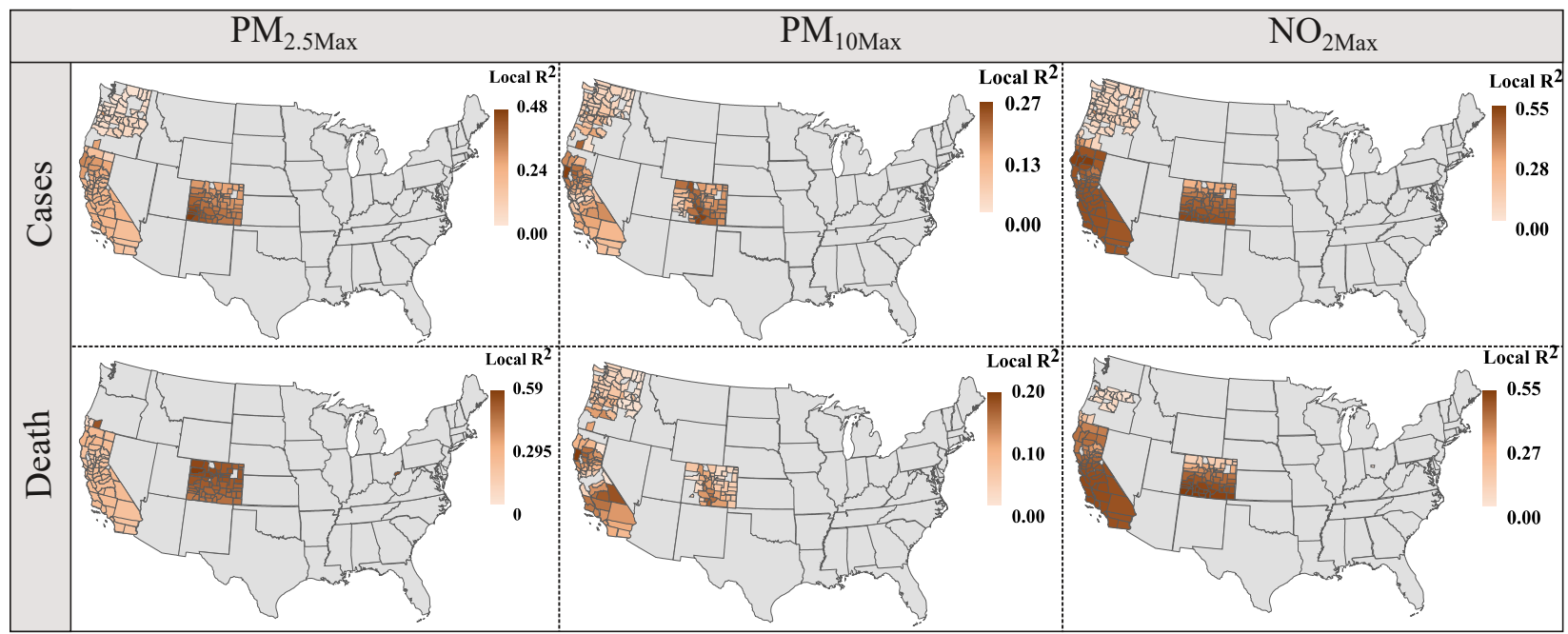

**Fig. 12** Spatial coefficient of determination values ($R^2$) exhibiting the spatial association between air pollution estimates and COVID incidents at local scale. $PM_{2.5Max}$, $PM_{10Max}$, $NO_{2Max}$ refers to the maximum average estimates of $PM_{2.5}$, $PM_{10}$, and $NO_2$ measured during the study period.

**Table. 1** Non parametric test to evaluate mean differences in $PM_{2.5}$, $PM_{10}$, and $NO_2$ concentration between fire (2020) and reference (2019) year.

| Month | Model | V | Expected value | Variance (V) | P-value (Two-tailed) | α |
|-------|-------|---|----------------|--------------|----------------------|---|
| August | Model 1 | 212 | 612.5 | 10106.25 | < 0.0001 | 0.05 |
| | Model 2 | 117 | 351.5 | 4393.75 | 0.000 | 0.05 |
| | Model 3 | 540 | 370.5 | 4754.75 | 0.014 | 0.05 |
| September | Model 4 | 416 | 612.5 | 10106.25 | 0.051 | 0.05 |
| | Model 5 | 168 | 351.5 | 4393.75 | 0.006 | 0.05 |
| | Model 6 | 621 | 370.5 | 4754.75 | 0.000 | 0.05 |
| October | Model 7 | 94 | 612.5 | 10106.25 | < 0.0001 | 0.05 |
| | Model 8 | 87 | 351.5 | 4393.75 | < 0.0001 | 0.05 |
| | Model 9 | 438 | 370.5 | 4754.75 | 0.331 | 0.05 |

**Table 2** Sperman correlation matrix showing the linear association between the air pollution and COVID incidents in different months. Bold values are statistically significant at different probability level.

| Month | N | Model | Max | | Mean | | Sum | |
|-------|---|-------|-----|---------|------|---------|-----|---------|
| | | | r | P value | r | P value | r | P value |
| August | 196 | $PM_{2.5}$ & Cases | **0.21** | **0.00** | **0.16** | **0.02** | 0.14 | 0.06 |
| | | $PM_{10}$ & Cases | **0.32** | **0.00** | **0.27** | **0.00** | **0.17** | **0.02** |
| | | $NO_2$ & Cases | 0.09 | 0.22 | -0.03 | 0.68 | 0.09 | 0.22 |
| September | | $PM_{2.5}$ & Cases | **0.25** | **0.00** | **0.22** | **0.00** | **0.18** | **0.01** |
| | | $PM_{10}$ & Cases | **0.31** | **0.00** | **0.26** | **0.00** | **0.18** | **0.01** |
| | | $NO_2$ & Cases | **0.22** | **0.00** | 0.13 | 0.08 | 0.13 | 0.06 |
| October | | $PM_{2.5}$ & Cases | **0.32** | **0.00** | **0.28** | **0.00** | **0.17** | **0.02** |
| | | $PM_{10}$ & Cases | **0.33** | **0.00** | **0.34** | **0.00** | **0.18** | **0.01** |
| | | $NO_2$ & Cases | **0.32** | **0.00** | **0.20** | **0.00** | **0.15** | **0.03** |
| August | | $PM_{2.5}$ & Death | **0.19** | **0.01** | 0.14 | 0.05 | 0.11 | 0.11 |
| | | $PM_{10}$ & Death | **0.31** | **0.00** | **0.27** | **0.00** | 0.14 | 0.05 |
| | | $NO_2$ & Death | 0.11 | 0.13 | 0.00 | 0.95 | 0.08 | 0.29 |
| September | | $PM_{2.5}$ & Death | **0.22** | **0.00** | **0.18** | **0.01** | **0.19** | **0.01** |
| | | $PM_{10}$ & Death | **0.30** | **0.00** | **0.24** | **0.00** | **0.21** | **0.00** |
| | | $NO_2$ & Death | **0.21** | **0.00** | 0.11 | 0.11 | **0.17** | **0.02** |
| October | | $PM_{2.5}$ & Death | **0.28** | **0.00** | **0.24** | **0.00** | **0.15** | **0.03** |
| | | $PM_{10}$ & Death | **0.28** | **0.00** | **0.30** | **0.00** | **0.16** | **0.02** |
| | | $NO_2$ & Death | **0.27** | **0.00** | **0.19** | **0.01** | **0.16** | **0.02** |

*Values in bold are different from 0 with a significance level alpha=0.05*

**Table. 3** Spatial regression estimates derived from MGWR model. A total of 18 models were developed for both cases and death factors.

| Month | Models | $R^2$ | Adj. $R^2$ | AIC | AICc | BIC | Adj.t-value (95%) |
|---|---|---|---|---|---|---|---|
| | | | | **Cases** | | | |
| August | $PM_{2.5}$&Cases | 0.348 | 0.308 | 497.192 | 499.025 | 537.981 | 2.067 |
| | $PM_{10}$&Cases | 0.336 | 0.294 | 501.132 | 502.987 | 542.174 | 2.125 |
| | $NO_2$&Cases | 0.428 | 0.404 | 464.078 | 464.968 | 492.349 | 2.527 |
| September | $PM_{2.5}$&Cases | 0.33 | 0.288 | 502.922 | 504.816 | 544.398 | 2.157 |
| | $PM_{10}$&Cases | 0.326 | 0.284 | 503.741 | 505.586 | 544.671 | 2.087 |
| | $NO_2$&Cases | 0.428 | 0.404 | 464.168 | 465.07 | 492.617 | 2.544 |
| October | $PM_{2.5}$&Cases | 0.257 | 0.211 | 522.761 | 524.573 | 563.323 | 2.03 |
| | $PM_{10}$&Cases | 0.254 | 0.208 | 523.633 | 525.476 | 564.54 | 2.088 |
| | $NO_2$&Cases | 0.322 | 0.297 | 496.051 | 496.808 | 522.051 | 2.563 |
| | | | | **Death** | | | |
| August | $PM_{2.5}$&Death | 0.314 | 0.272 | 506.949 | 508.722 | 547.059 | 2.006 |
| | $PM_{10}$&Death | 0.306 | 0.262 | 509.7 | 511.555 | 550.739 | 2.126 |
| | $NO_2$&Death | 0.401 | 0.377 | 472.848 | 473.718 | 500.788 | 2.528 |
| September | $PM_{2.5}$&Death | 0.255 | 0.228 | 514.256 | 515.011 | 540.227 | 2.224 |
| | $PM_{10}$&Death | 0.248 | 0.222 | 515.462 | 516.156 | 540.316 | 2.103 |
| | $NO_2$&Death | 0.461 | 0.438 | 452.571 | 453.469 | 480.958 | 2.544 |
| October | $PM_{2.5}$&Death | 0.291 | 0.247 | 513.541 | 515.354 | 554.117 | 2.033 |
| | $PM_{10}$&Death | 0.287 | 0.243 | 514.742 | 516.585 | 555.649 | 2.088 |
| | $NO_2$&Death | 0.428 | 0.407 | 462.975 | 463.78 | 489.815 | 2.561 |

**Table. 4** Spatial regression estimates derived from the MGWR model showing local association between the average concentration of pollutants during August to October 2020 and COVID cases and death.

| Models | $R^2$ | Adj. $R^2$ | AIC | AICc | BIC | Adj.t-value (95%) |
|---|---|---|---|---|---|---|
| | | | **Cases** | | | |
| $PM_{2.5}$Max&Cases | 0.315 | 0.272 | 507.041 | 508.87 | 547.793 | 2.039 |
| $PM_{2.5}$Mean&Cases | 0.32 | 0.278 | 505.489 | 507.304 | 546.077 | 2.044 |
| $PM_{10}$Max&Cases | 0.318 | 0.275 | 506.058 | 507.87 | 546.624 | 2.025 |
| $PM_{10}$Mean&Cases | 0.313 | 0.27 | 507.599 | 509.449 | 548.59 | 2.094 |
| $NO_2$Max&Cases | 0.542 | 0.525 | 418.836 | 419.581 | 444.619 | 2.489 |
| $NO_2$Mean&Cases | 0.409 | 0.385 | 470.152 | 471.012 | 497.921 | 2.543 |
| | | | **Death** | | | |
| $PM_{2.5}$Max&Death | 0.315 | 0.272 | 506.897 | 508.726 | 547.649 | 2.039 |
| $PM_{2.5}$Mean&Death | 0.318 | 0.276 | 506.007 | 507.821 | 546.595 | 2.044 |
| $PM_{10}$Max&Death | 0.319 | 0.277 | 505.607 | 507.42 | 546.173 | 2.026 |
| $PM_{10}$Mean&Death | 0.314 | 0.271 | 507.477 | 509.327 | 548.468 | 2.094 |
| $NO_2$Max&Death | 0.556 | 0.54 | 412.673 | 413.4 | 438.141 | 2.49 |
| $NO_2$Mean&Death | 0.443 | 0.421 | 458.315 | 459.175 | 486.084 | 2.543 |

Table 5 Summary table showing the reviewed studies that examined the association between air pollution (PM$_{2.5}$, PM$_{10}$ and NO$_2$) and COVID-19 cases/deaths.

| Pollutants | Author | Study area | Time period | Method used | Main findings |
|---|---|---|---|---|---|
| PM$_{2.5}$ | Zhu et al. 2020 | 120 cities in China | January 23 to February 29, 2020 | Generalized Additive Model (GAM) | 10 ug/m$^3$ increase in PM$_{2.5}$ was associated with a 2.24% increase in daily COVID-19 confirmed cases |
| | Fattorini & Regoli (2020) | 71 Italian province | February 24 to April 27, 2020 | Pearson correlation and regression analysis | R$^2$ = 0.340 (p < 0.01) with total confirmed cases |
| | Yao et al. (2020) | Wuhan | January 19 to March 15, 2020 | Time series analysis | CFR of COVID-19 increased by 0.86% (0.50%–1.22%) each 10 µg/m$^3$ increase in PM$_{2.5}$ |
| | Li et al. (2020) | Wuhan and XiaoGan | January 26 to February 29 2020 | Simple linear regression | R$^2$ = 0.174 (Wuhan), R$^2$ = 0.23 (XiaoGan) with daily confirmed cases |
| | Yao et al. (2020) | 49 cities of China | Upto March 22 2020 | Multiple linear regression | 10 ug/m$^3$ increase in PM$_{2.5}$ was associated with a 0.24% (0.01% - 0.48%) increase in daily COVID-19 fatality rate |
| | Zoran et al. (2020a) | Milan (Italy) | January 01 to April 30 | Pearson coefficient correlation | r = -0.39; r = 0.25; r = -0.53 for total cases, daily confirmed cases, and total deaths |
| | Frontera et al. (2020) | Italian regions | Upto March 31 2020 | Pearson correlation and regression analysis | R$^2$ = 0.64; p < 0.01 with total confirmed cases and R$^2$ = 0.53; p < 0.05 with deaths |
| | Wu et al. (2020) | 3000 counties in the U.S.A. | Upto April 04, 2020 | Zero-inflated negative binomial models | 1 ug/m$^3$ long-term exposure increase in PM$_{2.5}$ was associated with a 15% increase in COVID-19 death rate |
| | Adhikari and Yin (2020) | Queens county, New York, USA | March 01 to April 20, 2020 | Negative binomial regression model | Coefficient of estimates - 0.4029 for daily confirmed cases and -0.1151 for total death |
| | Vasquez-Apestegui et al (2020) | 24 districts of Lima, Perù | Upto June 12, 2020 | Multivariate regression model | Crude coefficient = 0.083, p < 0.05 (for total confirmed cases); Crude coefficient = 0.0016, p < 0.01 (for death); Crude coefficient = -0.014, p > 0.05 (for case fatality rate) |
| | Bashir et al. (2020) | California, USA | March 04 to April 24, 2020 | Spearman and Kendall correlation | Kendall r (-0.359); Spearman r (-0.453) (for confirmed cases); Kendall r (-0.339); Spearman r (-0.429) (for death); |
| | Travaglio et al., (2020) | England | February 1 and April 8, 2020 | generalised linear models, negative binomial regression | an increase of 1 m$^3$ in the long-term average of PM$_{2.5}$ was associated with a 12% increase in COVID-19 cases. |
| | Magazzino et al. (2020) | Paris, Lyon, and Marseille, Paris | March 18 to April 27, 2020 | Artificial Neural Networks (ANNs) | found new threshold levels of PM$_{2.5}$ for COVID-19: 17.4 µg/m$^3$ (PM$_{2.5}$) for Paris, 15.6 µg/m$^3$ (PM$_{2.5}$) for for Lyon;14.3 µg/m$^3$ (PM$_{2.5}$) for Marseille. Marseille, an increase in PM$_{2.5}$ concentrations above 14.3 µg/m$^3$ would generate a 79.01% increase in mortality |
| | Edgar and Hernández, 2020 | Victoria, Mexico | February 16 to June 06, 2020 | Pearson correlation analysis | Pearson r = 0.77 (last four weeks of the partial lockdown) and 0.64 ( twelve weeks of the partial lockdown) with total COVID-19 confirmed cases. |
| | Wu et al. (2020) | 3089 counties in the United States | Up to 18 June 2020 | Negative binomial mixed model | 1 µg/m$^3$ in the long-term average PM$_{2.5}$ is associated with a statistically significant 11% (95% CI, 6 to 17%) increase in the county's COVID-19 mortality rate. |
| | *Pozzer et al. (2020)* | Global | 2019 | Global atmospheric chemistry general circulation model (EMAC) | Globally, $PM_{2.5}$ contributed to 15% (95% CI 7–33%) COVID-19 mortality, 27% (CI 13 – 46%) in East Asia, 19% (CI 8– 41%) in Europe, and 17% (CI 6–39%) in North America. |
| | *Yihan Wu et al., (2020)* | 326 prefectures in mainland China | Up to April 21, 2020 | Negative binomial regression, Spearman's rank correlation | 1 µg m$^3$ increase in $PM_{2.5}$ can result in 1.95% (95% CI: 0.83–3.08%) rise of COVID-19 morbidity. Spearman's r = 0.35 (for COVID-19 morbidity counts). |
| | *This study* | 196 County of California, Colorado, Oregon, and Washington, USA | August 01 to October 31, 2020 | Spearman correlation, Ordinary Least Square Regression, Multiscale Geographically weighted regression, Negative binomial regression | Spearman's r = 0.26 (for daily confirmed cases) and r = 0.23 (for death) |
| **PM$_{10}$** | *Zhu et al. 2020* | 120 cities in China | January 23 to February 29, 2020 | Generalized Additive Model (GAM) | 10 ug/m$^3$ increase in $PM_{10}$ was associated with a 1.76% increase in daily COVID-19 confirmed cases |
| | *Coccia (2020)* | 55 Italian province capitals | 17th March 2020 to 7th April 2020 | Hierarchical multiple regression model | Cities with having more than 100 days of air pollution (exceeds the limits set for $PM_{10}$) have a very high average number of infected people (about 3350) |
| | *Fattorini & Regoli (2020)* | 62 Italian province | February 24 to April 27, 2020 | Pearson correlation and regression analysis | $R^2 = 0.267$ (p < 0.01) with total confirmed cases |
| | *Yao et al. (2020)* | Wuhan | January 19 to March 15, 2020 | Time series analysis | Fatality rate of COVID-19 increased by 0.83% (0.49%–1.17%) for each 10 µg/m$^3$ in $PM_{10}$ |
| | *Li et al. (2020)* | Wuhan and XiaoGan | January 26 to February 29 2020 | Simple linear regression | $R^2 = 0.105$ (Wuhan), $R^2 = 0.158$ (XiaoGan) with daily confirmed cases |
| | *Yao et al. (2020)* | 49 cities of China | Upto March 22 2020 | Multiple linear regression | 10 ug/m$^3$ increase in $PM_{10}$ was associated with a 0.26% (0.00% - 0.51%) increase in daily COVID-19 fatality rate |
| | *Zoran et al. (2020a)* | Milan (Italy) | January 01 to April 30 | Pearson coefficient correlation | r = -0.30; r = 0.35; r = -0.49 for total cases, daily confirmed cases, and total deaths |
| | *Bashir et al. (2020)* | California, USA | March 04 to April 24, 2020 | Spearman and Kendall correlation | Kendall r (-0.287); Spearman r (-0.375) (for confirmed cases); Kendall r (-0.267); Spearman r (-0.350) (for death); |
| | *Magazzino et al. (2020)* | Paris, Lyon, and Marseille, Paris | March 18 to April 27, 2020 | Artificial Neural Networks (ANNs) | found new threshold levels of $PM_{10}$ for COVID-19: 29.6 µg/m$^3$ ($PM_{10}$) for Paris, 20.6 µg/m$^3$ ($PM_{10}$) for Lyon;22.04 µg/m$^3$ ($PM_{10}$) for Marseille. In the city of Paris, an increase in $PM_{10}$ concentration beyond the 29.6 µg/m$^3$ threshold could generate a 63.2% increase in mortality (in a COVID-19 pandemic). For Lyon, any value above 20.6 µg/m$^3$ in $PM_{10}$ would generate an increase in deaths of 56.12%. |
| | *Edgar and Hernández, 2020* | Victoria, Mexico | February 16 to June 06, 2020 | Pearson correlation analysis | Pearson r = 0.79 (last four weeks of the partial lockdown) and 0.69 ( twelve weeks of the partial lockdown) with total COVID-19 confirmed cases. |

| | | | | |
|---|---|---|---|---|
| | *Marquès et al., (2020)* | Tarragona Province (Catalonia, Spain) | March 8, 2020, and May 10, 2020 | Pearson correlation analysis | $R^2 = 0.11$ (Chronic exposure of $PM_{10}$ (2014 - 2019) and $R^2 = 0.01$ (Outbreak exposure of $PM_{10}$ (2020) with confirmed cases per 1000 persons |
| | *Yihan Wu et al., (2020)* | 326 prefectures in mainland China | Up to April 21, 2020 | Negative binomial regression, Spearman's rank correlation | $1 \ \mu g \ m^3$ increase of $PM_{10}$ can result in 0.55% (95% CI: –0.05–1.17%), rise of COVID-19 morbidity. Spearman's r = 0.15 (for COVID-19 morbidity counts. |
| | *This study* | 196 County of California, Colorado, Oregon, and Washington, USA | August 01 to October 31, 2020 | Spearman correlation, Ordinary Least Square Regression, Multiscale Geographically weighted regression, Negative binomial regression | Spearman's r = 0.32 (for daily confirmed cases) and r = 0.30 (for death) |
| **$NO_2$** | *Zhu et al. 2020* | 120 cities in China | January 23 to February 29, 2020 | Generalized Additive Model (GAM) | 10 $ug/m^3$ increase in $NO_2$ was associated with a 6.94% increase in daily COVID-19 confirmed cases |
| | *Fattorini & Regoli (2020)* | 62 Italian province | February 24 to April 27, 2020 | Pearson correlation and regression analysis | $R^2 = 0.247$ ($p < 0.01$) with total confirmed cases |
| | *Li et al. (2020)* | Wuhan and XiaoGan | January 26 to February 29 2020 | Simple linear regression | $R^2 = 0.329$ (Wuhan), $R^2 = 0.158$ (XiaoGan) with daily confirmed cases |
| | *Ogen (2020)* | 66 administrative regions in Italy, Spain, France, Germany | Upto end of February 2020 | Descriptive analysis | 83% of COVID-19 fatality in the study regions are associated with $NO_2 > 100 \ \mu mol/m^2$ range |
| | *Zoran et al. (2020a)* | Milan (Italy) | January 01 to April 30 | Pearson coefficient correlation | r = -0.55; r = -0.35; r = -0.58 for total cases, daily confirmed cases, and total deaths |
| | *Bashir et al. (2020)* | California, USA | March 04 to April 24, 2020 | Spearman and Kendall correlation | Kendall r (-0.514); Spearman r (-0.736) (for confirmed cases); Kendall r (-0.485); Spearman r (-0.731) (for death); |
| | *Huang and Brown (2020)* | 401 counties of Germany | Upto 13th, September, 2020 | Poisson log-linear model | $1 \ \mu g \ m^3$ increase in long-term exposure to $NO_2$ increasing the COVID-19 incidence rate by 5.58% |
| | *Marquès et al., (2020)* | Tarragona Province (Catalonia, Spain) | March 8, 2020, and May 10, 2020 | Pearson correlation analysis | $R^2 = 0.55$ (Chronic exposure of $NO_2$ (2014 - 2019) and $R^2 = 0.59$ (Outbreak exposure of $NO^2$ (2020) with confirmed cases per 1000 persons |
| | *Yihan Wu et al., (2020)* | 326 prefectures in mainland China | Up to April 21, 2020 | Negative binomial regression, Spearman's rank correlation | $1 \ \mu g \ m^3$ increase of $NO_2$ can result in 4.63% (95% CI: 3.07–6.22%) rise of COVID-19 morbidity. Spearman's r = 0.37 (for COVID-19 morbidity counts. |
| | *This study* | 196 County of California, Colorado, Oregon, and Washington, USA | August 01 to October 31, 2020 | Spearman correlation, Ordinary Least Square Regression, Multiscale Geographically weighted regression, Negative binomial regression | Spearman's r = 0.21 (for daily confirmed cases) and r = 0.20 (for death) |